\documentclass[preprint]{JHEP3} 

\usepackage{epsfig,multicol}

\def\fr#1#2{\hbox{${#1\over #2}$}}

\def\+{{(+)}}  \def\-{ {(-)} }   \def\0{ {(0)} }
\def\1{ {(1)} }  \def\2{ {(2)} }

\def\con{\omega}

\def\sq{Q\kern-6pt/}
\def\sQ{Q\kern-12pt\nearrow}

\def\be{\begin{equation}}             \def\ee{\end{equation}}
\def\ba{\begin{array}{rcl}}           \def\ea{\end{array}}
\def\beqa{\begin{eqnarray} }          \def\eeqa{\end{eqnarray} }
\def\beqalign{\begin{eqalign}}        \def\eeqalign{\end{eqalign}}
\def\leq#1{\label{eq:#1}}             \def\eq#1{(\ref{eq:#1})}
\def\bsubeq{\begin{subequations}}     \def\esubeq{\end{subequations}}
\def\bitem{\begin{itemize}}           \def\eitem{\end{itemize}}

\newcommand\fverb{\setbox\pippobox=\hbox\bgroup\verb}
\newcommand\fverbdo{\egroup\medskip\noindent%
            \fbox{\unhbox\pippobox}\ }
\newcommand\fverbit{\egroup\item[\fbox{\unhbox\pippobox}]}
\newbox\pippobox

\title{Torsion and nonmetricity in the stringy geometry}

\author{B. Sazdovi\'c \thanks{Work supported in part by the Serbian Ministry of Science, Technology
and Development under contract No. 1486.}~ \\
  Institute of Physics, 11001 Belgrade, P.O.Box 57, Serbia   \\
    E-mail: \email{sazdovic@phy.bg.ac.yu}}
\received{February 20, 2001}        
\accepted{November 27, 2001}        

\abstract{In the present article, we study the space-time geometry felt by
probe bosonic string moving in antisymmetric and dilaton background fields.
This space-time geometry we shall call the stringy geometry. In particular,
the presence of the antisymmetric field leads to the space-time torsion, and
the presence of the dilaton field leads to the space-time nonmetricity. We
generalize the geometry of surfaces embedded in space-time to the case when
torsion and nonmetricity are present. We define the mean extrinsic curvature
for Minkowski signature and introduce the concept of mean torsion. Its
orthogonal projection defines the dual mean extrinsic curvature. In this
language, one field equation is just the equality of mean extrinsic curvature
and dual mean extrinsic curvature, which we call C-duality relation. In the
torsion and nonmetricity free case, the world-sheet is a minimal surface,
specified by the requirement that mean extrinsic curvature vanishes.
Generally, it is stringy C-dual surface. In the presence
of the dilaton field, which breaks conformal invariance, the conformal factor
which connects intrinsic and induced metrics, is determined as a function of
the dilaton field itself. We also derive the integration measure for the
space-time with stringy nonmetricity.}


\begin{document}


\section{Introduction}

In the paper \cite{BS}, we investigated classical dynamics of the bosonic
string in the background metric, antisymmetric and dilaton fields. The
starting Lagrangian has been used in the literature  \cite{FT,CFMP, BNS}, in
order to study the two-dimensional conformal invariance on the quantum level.
In those papers the classical space-time equations of motion have been
obtained from the world-sheet quantum conformal invariance.

In the ref. \cite{BS}, we made classical canonical nonperturbative
investigations, treating world-sheet fields as variables in the theory and
space-time fields as a background depending on the coordinates $x^\mu$. We
obtained the nonlinear realization of the Virasoro generators. We also derived
the Hamiltonian  equations of motion and shortly discussed the target space
torsion and nonmetricity, recognized by the string.

In the present paper, we are going to investigate in detail the space-time
geometry felt by the string, and offer geometrical interpretation of the
equations of motion. In Sec. 2, we will consider general theory of surfaces
embedded in space-time with torsion and nonmetricity, and in Secs. 3 and 4
we will apply this results to the bosonic string theory in the background
fields.

Starting with the known rules of the space-time parallel transport, in Sec. 2
we introduce the torsion and nonmetricity. We decompose the arbitrary
connection in terms of the Christoffel one, contortion and nonmetricity. Then
we define the induced world-sheet variables: metric tensor and connection, and
extrinsic one: second fundamental form (SFF). The world-sheet tangent vector,
after parallel transport along world-sheet line with space-time connection is
not necessarily a tangent vector. Its world-sheet projection defines the
induced connection and its normal projection defines the SFF. When the metric
postulate is not valid, there are two forms of the world-sheet connections and
two forms of SFF. The two forms, of both the connection and the SFF, differ by
terms proportional to the nonmetricity. We also introduce the induced torsion
and induced nonmetricity and find the relations between space-time and
world-sheet covariant derivatives. We generalized the geometrical
interpretation of the SFF to the case where the space-time has nontrivial
torsion and nonmetricity. In the standard approach the torsion does not
contribute to the mean extrinsic curvature (MEC) ${}^\circ H_i$. We define
dual mean extrinsic curvature (DMEC) ${}^\ast H_i$ as an orthogonal projection
of the mean torsion. This enables us to define the self-dual (self-antidual)
condition, under C-duality ${}^\circ H_i = \pm {}^\ast H_i$.

In Sec. 3, we formulate the bosonic string theory and shortly repeat some
results of ref. \cite{BS}, such as canonical derivation of the equations of
motion. We independently derive the Lagrangian equations of motion and connect
them to the previous ones. We introduce the stringy space-time felt by the
probe string and find the expressions for stringy torsion and stringy
nonmetricity.

In Sec. 4, we investigate contributions of the background fields to the
space-time geometry. The equations of motion with respect to the intrinsic
world-sheet metric relate this intrinsic metric with the induced one. In the
absence of the dilaton field $\Phi$, the theory is conformally  invariant, so
that the intrinsic metric tensor is equal to the induced one up to the
conformal factor $\lambda$. The presence of the dilaton field breaks the
conformal invariance and determines this conformal factor.

Equations of motion, obtained by variation with respect to $x^\mu$, have $D$
components. Two of them determine the contracted intrinsic connection in terms
of the corresponding induced expression. The other $D-2$ are of the form
${}^\star H_i = \pm {}^\ast H_{\mp i}$, where ${}^\star H_i$ is stringy MEC
and ${}^\ast H_{\pm i}$ are the two forms of DMEC. They define world-sheet as
a stringy C-dual (antidual) surface. In two particular cases, ---the vanishing
torsion and the vanishing nonmetricity--- the field equations turn to the
equations of stringy minimal world-sheet ${}^\star H_i = 0$ and C-dual
(antidual) world-sheet $H_i = \pm {}^\ast H_{\mp i}$, respectively. In the
case of Riemann space-time, when both torsion and nonmetricity vanish, they
turn to the equations of minimal world-sheet $H_i = 0$.

The presence of the space-time metric $G_{\mu \nu}$ produces just the standard
Christoffel connection. In this case, the string can see the space-time as
Riemannian one. The field strength of the antisymmetric tensor contributes to
the space-time torsion. So, string with background fields $G_{\mu \nu}$ and
$B_{\mu \nu}$ feels Riemann-Cartan space-time. Finally, the dilaton field
$\Phi$ produces the nonmetricity of the space-time. The target space observed
by the string, when all three background fields $G_{\mu \nu}$, $B_{\mu \nu}$
and $\Phi$ are present, we will call the stringy space-time.

In sec. 5, we consider the possible forms of the space-time action. We obtain
the new integration measure for spaces with nonmetricity from the requirements
that the measure is preserved under parallel transport and that it enables
integration by parts. We discuss how our action is related to the action of
the papers \cite{CFMP,BNS}.

Appendix A is devoted to the world-sheet geometry, and Appendix B to the
classification of space-time geometry and to the world-sheet as an embedded surface.

\section{Geometry of surfaces embedded in space-time with torsion and nonmetricity}

The geometry of surfaces, when the world-sheet is embedded in
curved space-time, has been investigated in the literature, see
\cite{BN,Sor}. In this section we will generalize these results for
the space-times with nontrivial torsion and nonmetricity. For some
details of the subsection 2.1 see refs. \cite{MB,HMMN}.

\subsection{Geometry of space-time with torsion and nonmetricity}

In the curved spaces, the operations on tensors are covariant only if they are
realized in the same point. In order to compare the vectors from different
points we need the rule for parallel transport. The parallel transport of the
vector $V^\mu (x)$, from the point $x$ to the point $x+dx$, produce the vector
${}^\circ V^\mu_\parallel = V^\mu + {}^\circ \delta V^\mu$, where
\be
{}^\circ \delta V^\mu = - {}^\circ \Gamma_{\rho \sigma}^\mu V^\rho d x^\sigma  \,  .  \leq{ptcn}
\ee
The variable ${}^\circ \Gamma_{\rho \sigma}^\mu$ is the {\bf affine linear connection}.
The rule for the parallel transport of the covector $U_\mu(x)$,
can be obtained from the requirement that the invariant vector product is not changed
under parallel transport ${}^\circ \delta (V^\mu U_\mu)=0$, so that
\be
{}^\circ \delta U_\mu =  {}^\circ \Gamma_{\mu \nu }^\rho U_\rho d x^\nu  \,  .
\ee
Now, we are able to compare vectors from the points $x$ and $x+dx$, and define the covariant derivative
\be
{}^\circ D V^\mu = V^\mu (x+dx) - {}^\circ V^\mu_\parallel = d V^\mu - {}^\circ \delta V^\mu =
(\partial_\nu V^\mu + {}^\circ \Gamma_{\rho \nu}^\mu V^\rho ) d x^\nu \equiv  {}^\circ D_\nu V^\mu d x^\nu     \,   .
\ee

Let us define the {\bf geodesic line} as a self-parallel line,
which means that the tangent vector $ t^\mu = {d x^\mu \over d s} = \dot{x}^\mu$, stay parallel
after parallel transport $\, t^\mu + {}^\circ t^\mu = t^\mu (s+ds)$. It produce the equation
\be
\ddot{x}^\mu + {}^\circ \Gamma^\mu_{\rho \sigma} \dot{x}^\rho
\dot{x}^\sigma= 0 \,   ,
\ee
stating that the covariant derivative along the geodesic is zero
$\dot{x}^\nu \, {}^\circ \! D _\nu \dot{x}^\mu = 0$.

The connection is not necessary symmetric in the lower indices, and its
antisymmetric part is the {\bf torsion}
\be
{}^\circ T^\rho_{\mu \nu} ={}^\circ \Gamma^\rho_{\mu \nu} - {}^\circ \Gamma^\rho_{\nu \mu} \,  .
\ee
It has a simple geometrical interpretation. Consider two geodesics $\ell_1$
and $\ell_2$, starting at the point $A$, with the corresponding unit tangent
vectors $ t^\mu_r = {d x^\mu \over d s_r}  \,(r=1,2)$, (Fig. 1). Let us
perform parallel transport of the vector $t^\mu_2$, along geodesic $\ell_1$,
to the point $B$ at the distance $d s_1 = d \ell_1$. The final vector we
denote by $\tilde{t}^\mu_2$. It defines direction of the geodesic
$\tilde{\ell}_2$, which starts at the point $B$ and ends at the point $D_1$,
at the distance $d \tilde{s}_2 = d \ell_2$. Similarly, we perform parallel
transport of the vector $t^\mu_1$, along geodesic $\ell_2$, to the point $C$
at the distance $d s_2 = d \ell_2$. We obtain the vector $\tilde{t}^\mu_1$,
which determines the geodesic $\tilde{\ell}_1$. The point $D_2$, lies on the
geodesic $\tilde{\ell}_1$ at the distance $d \tilde{s}_1 = d \ell_1$ from
point $C$.
\smallskip
\FIGURE{\epsfig{file=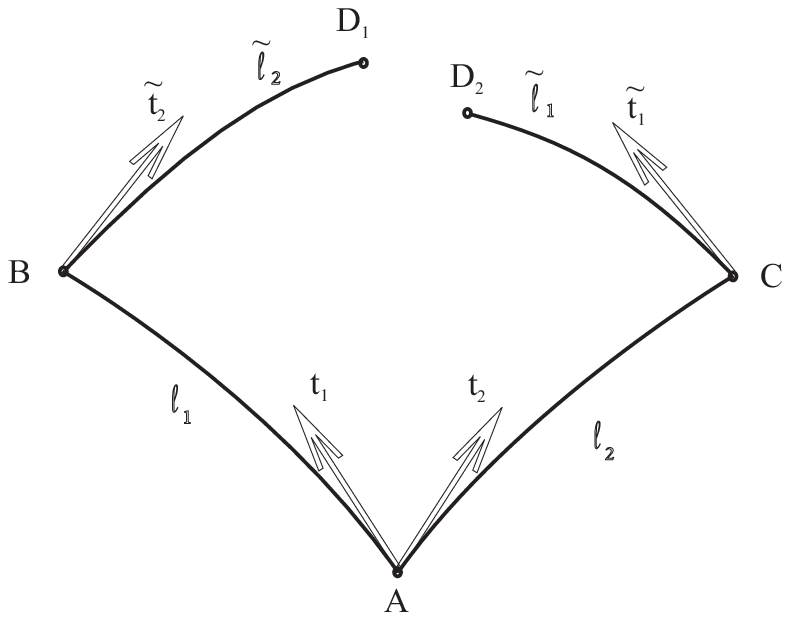,width=9cm}
        \caption[Example]{{\it Geometrical meaning of the torsion.}}%
    \label{torsion}}

In the curved space-time, this figure is not necessary closed. The difference
of the coordinates at the points $D_2$ and $D_1$
is proportional to the torsion
\be
x^\mu(D_2) - x^\mu(D_1)  = {}^\circ T^\mu{}_{\rho \sigma}\, t_1^\rho \, t_2^\sigma \, d \ell_1 \, d \ell_2    \,   .       \leq{cod}
\ee
In fact {\it the torsion measures the non-closure of the "rectangle" $ABCD$}.

The {\bf metric tensor} $G_{\mu \nu}$ is a new variable, independent on the connection.
This is a non-degenerate symmetric tensor, which enables us to calculate scalar product
$VU= G_{\mu \nu} V^\mu U^\nu$, in order to measure lengths and angles.

We already learned, that covariant derivative is responsible for the
comparison of the vectors from different points. What variable is responsible
for comparison of the lengths of the vectors? The square of the length of the
vector $V^\mu(x)$ is $V^2(x)= G_{\mu \nu}(x) V^\mu(x) V^\nu (x)$. The square of
the length of its parallel transport to the point $x+dx$, is  ${}^\circ
V_\parallel^2(x+dx)= G_{\mu \nu}(x+dx){}^\circ V^\mu_\parallel \, {}^\circ
V^\nu_\parallel$. Let us stress, that we used the local metric tensors in both
expressions and realized the parallel transport with the connection ${}^\circ
\Gamma^\rho_{\nu \mu}$. If we remember the invariance of the scalar product
under the parallel transport, than the difference of the squares of the
vectors is
\be
{}^\circ \delta V^2 =  {}^\circ V_\parallel^2 (x+dx) - V^2 (x)  =
[G_{\mu \nu}(x+dx) - G_{\mu \nu}(x) - {}^\circ \delta G_{\mu \nu}(x)]
\, {}^\circ V^\mu_\parallel \, \, {}^\circ V^\nu_\parallel    \,   .
\ee
Up to the higher terms we have
\be
{}^\circ \delta V^2 =  [d G_{\mu \nu}(x) - {}^\circ \delta G_{\mu \nu}(x)] V^\mu V^\nu =
 {}^\circ D G_{\mu \nu} V^\mu V^\nu  \equiv - d x^\rho \, {}^\circ\! Q_{\rho \mu \nu} V^\mu V^\nu \,  , \leq{len}
\ee
where we introduced the {\bf nonmetricity} as a covariant derivative of the metric tensor
\be
{}^\circ Q_{\mu \rho \sigma}=- {}^\circ D_\mu G_{\rho \sigma}  \,  .  \leq{nm}
\ee
Beside the lengths, the nonmetricity also changes the angle between the vectors
$V_1^\mu$ and $V_2^\mu$, according to the relation
\be
{}^\circ \delta \cos( \angle ( V_1 , V_2)) =  {1 \over 2 \sqrt{V_1^2 V_2^2 }}
\left[2 V_1^\rho V_2^\sigma - \left( {V_1^\rho V_1^\sigma \over V_1^2} + {V_2^\rho V_2^\sigma \over V_2^2} \right) (V_1 V_2) \right]
{}^\circ  Q_{\mu \rho \sigma} d x^\mu       \,  .    \leq{ang}
\ee

Note that we performed the parallel transport of the vectors, but not of the
metric tensor. It means that for length calculation in the point $x +dx$, we
used the metric tensor $G_{\mu \nu}(x + dx)$, which lives in the same point,
and not the tensor $G_{\mu \nu} + {}^\circ \delta G_{\mu \nu}$ obtained after
parallel transport from the point $x$. The requirement for the equality of
these two tensors is known in the literature as a metric postulate. In fact,
it is just some kind of compatibility between the metric and connection, such
that metric after parallel transport is equal to the local metric. Here we
will not accept this requirement, because the difference of these two tensors
is the origin of the nonmetricity. So, {\it the nonmetricity measures the
deformation of lengths and angles during the parallel transport}.

We also define the Weyl vector as
\be
{}^\circ Q_\mu  = {1 \over D} G^{\rho \sigma } {}^\circ Q_{\mu \rho \sigma}    \,    ,     \leq{wv}
\ee
where $D$ is the number of space-time dimensions. When the traceless part of
the nonmetricity vanishes
\be
{}^\circ \sQ_{\mu \rho \sigma} \equiv {}^\circ Q_{\mu \rho \sigma} - G_{\rho \sigma} {}^\circ Q_\mu =0     \,       ,   \leq{ap}
\ee
the parallel transport preserves the angles but not the lengths. Such geometry
is known as a Weyl geometry.

Following the paper \cite{HMMN}, we can decomposed the given connection
${}^\circ \Gamma^\mu_{\nu \rho}$ in terms of the Christoffel connection,
contortion and nonmetricity. If we introduce the Schouten braces according to
the relation
\be
\{ \mu \rho \sigma \} = \sigma \mu \rho + \rho \sigma \mu - \mu \rho \sigma  \,       ,
\ee
then the Christoffel connection \eq{coG}, can be expressed as $\Gamma_{\mu , \rho \sigma} =
\frac{1}{2} \partial_{ \{ \mu } G_{ \rho \sigma \} }$.
The contortion ${}^\circ K_{\mu \rho \sigma}$ is defined in terms of the torsion
\be
{}^\circ K_{\mu \rho \sigma} = \frac{1}{2} {}^\circ T_{ \{ \sigma \mu \rho \} } =
\frac{1}{2} ({}^\circ T_{\rho \sigma \mu} + {}^\circ T_{\mu \rho \sigma} - {}^\circ T_{\sigma \mu \rho})    \,  .  \leq{K}
\ee
By definition, the contortion is antisymmetric under first two indices ${}^\circ K_{\mu \rho
\sigma}= - {}^\circ K_{\rho \mu \sigma}$.

The Schouten braces of the nonmetricity can be solved in terms of the connection
\be
{}^\circ \Gamma_{\mu ,\rho \sigma} = \Gamma_{\mu ,\rho \sigma} + {}^\circ K_{\mu \rho \sigma} +
\frac{1}{2} {}^\circ Q_{\{ \mu \rho \sigma \} }   \,   .     \leq{cde}
\ee
The first term is the Christoffel connection , which depends on the metric but
which does not transforms as a tensor. The second one is the contortion \eq{K}
and the third one is Schouten braces of the nonmetricity \eq{nm}. The last two
terms transform as a tensors.

The first and third terms are symmetric in $\rho, \sigma$ indices. In the
second term we can separate symmetric and antisymmetric parts, ${}^\circ
K_{\mu \rho \sigma}= {}^\circ K_{\mu (\rho \sigma)}+ \frac{1}{2} {}^\circ
T_{\mu \rho \sigma}$, where he symmetric part of the arbitrary tensor $X_{\mu
\nu}$ we denote as $X_{(\mu \nu)} \equiv \fr{1}{2}(X_{\mu \nu} + X_{\nu
\mu})$. Consequently, we have
\be
{}^\circ \Gamma^\mu_{\rho \sigma} = {}^\circ \Gamma^\mu_{(\rho
\sigma)} + \frac{1}{2} {}^\circ T^\mu_{\rho \sigma} \,  .
\ee

\subsection{Induced and extrinsic geometry}

Let $x^\mu(\xi) \,  (\mu =0,1,...,D-1)$ be the coordinates of the $D$
dimensional space-time $M_D$ and $\xi^\alpha \, (\xi^0 =\tau , \xi^1=\sigma)$
the coordinates of two dimensional world-sheet $\Sigma$, spanned by the
string. The corresponding derivatives we will denote as $\partial_\mu \equiv
{\partial \over \partial x^\mu}$ and $\partial_\alpha \equiv {\partial \over
\partial \xi^\alpha}$. We will use the local space-time basis, relating with
the coordinate one by the vielbein $E^\mu_A = \{\partial_\alpha x^\mu ,
n_i^\mu \}$. Here, $\partial_\alpha x^\mu = \{ {\dot x^\mu} , x'^\mu \}$ is
local world-sheet basis and $n^\mu_i \, (i=2,3,...,D-1)$ are local unit
vectors, normal to the world-sheet.

The two dimensional induced metric tensor is defined by the requirement, that
any world-sheet interval measured by the target space metric, has the same
length as measured by the induced one. For a given space-time metric tensor
$G_{\mu \nu}$, the world-sheet {\bf induced metric tensor} takes the form
\be
G_{\alpha \beta} = G_{\mu \nu} \partial_\alpha x^\mu
\partial_\beta x^\nu \,  .  \leq{imt}
\ee
Similarly, the induced metric of a $D-2$ dimensional space, normal to the
world-sheet, is $G_{i j} = G_{\mu \nu}  n_i^\mu n_j^\nu$. The mixed induced
metric tensor $G_{\alpha i} = G_{\mu \nu} \partial_\alpha x^\mu n_i^\nu$,
vanishes by definition.

The world-sheet projection and the orthogonal projection of the arbitrary
space-time covector $V_\mu$, we will denote as
\be
v_\alpha = \partial_\alpha x^\mu V_\mu \,  , \qquad
v_i = n_i^\mu V_\mu     \,   .      \leq{pr}
\ee
In the space-time basis, tangent and normal vectors to the world-sheet
$\Sigma$ can be expressed respectively as
\be
V^\mu_\Sigma =  \partial_\alpha x^\mu v^\alpha  \,   ,
\qquad   V^\mu_\perp =  n_i^\mu  v^i  \,   .
\ee

a) {\it Parallel transport of the world-sheet tangent covectors.} Let us
perform the parallel transport of the covector $V_\mu^\Sigma$ along
world-sheet line, from the point $\xi^\alpha$ to the point $\xi^\alpha + d
\xi^\alpha$, using the space-time connection ${}^\circ \Gamma^\nu_{\mu \rho}$
(Fig. 2). We obtain the covector
\be
{}^\circ V^\Sigma_{\parallel \mu} = V^\Sigma_\mu + {}^\circ \delta
V^\Sigma_\mu \, ,  \leq{ptv}
\ee
where
\be
{}^\circ \delta V_\mu^\Sigma ={}^\circ \Gamma^\nu_{\mu \rho} V_\nu^\Sigma d x^\rho ={}^\circ \Gamma^\nu_{\mu \rho}
V_\nu^\Sigma \partial_\gamma x^\rho d \xi^\gamma \,  .
\ee

In the local basis, at the point $\xi + d \xi$, its world-sheet projection
has the form
\be
{}^\circ v_{\parallel \alpha}^\Sigma  = \partial_\alpha x^\mu (\xi + d \xi) {}^\circ
V^\Sigma_{\parallel \mu}  \,  .
\ee

\FIGURE{\epsfig{file=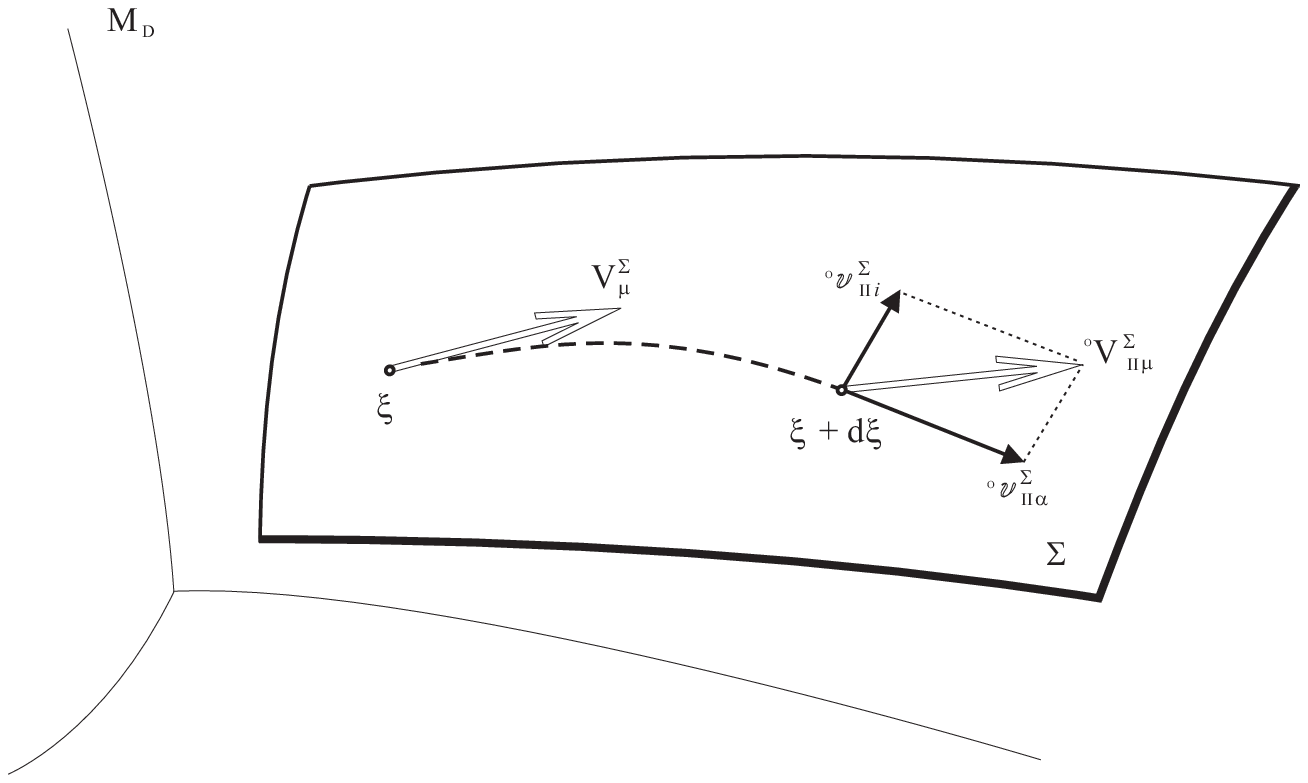,width=13cm}
        \caption[Example]{{\it Definition of the induced connection}
        ${}^\circ \Gamma^\beta_{\alpha \gamma }$  {\it and SFF}
        ${}^\circ {\bar b}_{i \alpha \beta}$
        {\it from the parallel transport of the world-sheet tangent covector.
        Here} ${}^\circ  v_{\parallel \alpha}^\Sigma \equiv v_\alpha  +
        {}^\circ \Gamma^\beta_{\alpha \gamma }
        v_\beta \, d \xi^\gamma \,$ {\it and} ${}^\circ v^\Sigma_{\parallel i} \equiv -
{}^\circ {\bar b}_{i \alpha \beta} \, v^\alpha d \xi^\beta $.}%
    \label{tan}}

Let us introduce the world-sheet {\bf induced connection} ${}^\circ \Gamma^\beta_{\alpha \gamma
}$. We demand that it defines the rule of the parallel transport of the
world-sheet covector $v_\alpha$, along the same world-sheet line, to
the projection ${}^\circ v_{\parallel \alpha}^\Sigma$. So, we have by
definition
\be
{}^\circ v_{\parallel \alpha}^\Sigma  = v_\alpha +{}^\circ \delta v_\alpha   \,  ,  \leq{ptd}
\ee
where
\be
{}^\circ \delta v_\alpha ={}^\circ \Gamma^\beta_{\alpha \gamma } v_\beta d \xi^\gamma \,  .  \leq{ptw}
\ee
It produce the expression for the induced connection
\be
{}^\circ \Gamma^\gamma_{\alpha \beta} = G^{\gamma \delta} \partial_\delta x^\mu G_{\mu \nu}
({}^\circ \Gamma_{\rho \sigma}^\nu   \partial_\alpha x^\rho \partial_\beta x^\sigma
+  \partial_\beta  \partial_\alpha x^\nu )=
G^{\gamma \delta} \partial_\delta x^\mu   G_{\mu \nu}
{}^\circ D_\beta \partial_\alpha x^\nu   \,  , \leq{icn}
\ee
where ${}^\circ D_\alpha V^\mu = \partial_\alpha x^\nu {}^\circ D_\nu V^\mu$
is space-time covariant derivative along world-sheet direction.

The orthogonal projection of the covector ${}^\circ V^\Sigma_{\parallel
\mu}\,$, defines the {\bf second fundamental form}, ${}^\circ {\bar b}_{i
\alpha \beta}\,$, trough the equation
\be
n_i^\mu (\xi+ d \xi) {}^\circ V^\Sigma_{\parallel \mu}
\equiv {}^\circ v^\Sigma_{\parallel i} = - {}^\circ {\bar b}_{i \alpha
\beta}v^\alpha d \xi^\beta  \,   ,
\ee
or explicitly
\be
{}^\circ {\bar b}_{i \alpha \beta} = n_i^\mu \, {}^\circ \! D_\beta ( G_{\mu \nu} \partial_\alpha x^\nu )
 = -  \partial_\alpha x^\nu G_{\mu \nu} {}^\circ \! D_\beta  n_i^\mu         \,  .    \leq{sff2}
\ee
The SFF define the extrinsic geometry.

b) {\it Parallel transport of the world-sheet orthogonal covectors.} If we
perform parallel transport of the covector $V_\mu^\bot$, orthogonal to the
world-sheet (Fig. 3), we obtain
\be
{}^\circ V^\bot_{\parallel \mu} = V^\bot_{\mu} + {}^\circ \delta V^\bot_{\mu}
\,  ,  \qquad  {}^\circ \delta V^\bot_{\mu}= {}^\circ \Gamma^\nu_{\mu \rho} V^\bot_{\nu} d x^\rho =
{}^\circ \Gamma^\nu_{\mu \rho} V^\bot_{\nu} \partial_\alpha x^\rho d
\xi^\alpha \,   .
\ee
Its normal projection, defines the induced connection of the $D-2$ dimensional
space, normal to the world-sheet
\be
{}^\circ \Gamma^i_{j \alpha } = G^{i k} n^\mu_k   G_{\mu \nu} {}^\circ D_\alpha n^\nu_j
\,  ,
\ee
and its world-sheet projection defines also the SFF, which we denote by
${}^\circ b_{i \alpha \beta}$
\be
\partial_\alpha x^\mu (\xi + d \xi) {}^\circ V^\bot_{\parallel
\mu} \equiv {}^\circ v^\bot_{\parallel \alpha}
= v^i \, {}^\circ b_{i \alpha \beta} d \xi^\beta  \,  .
\ee

\FIGURE{\epsfig{file=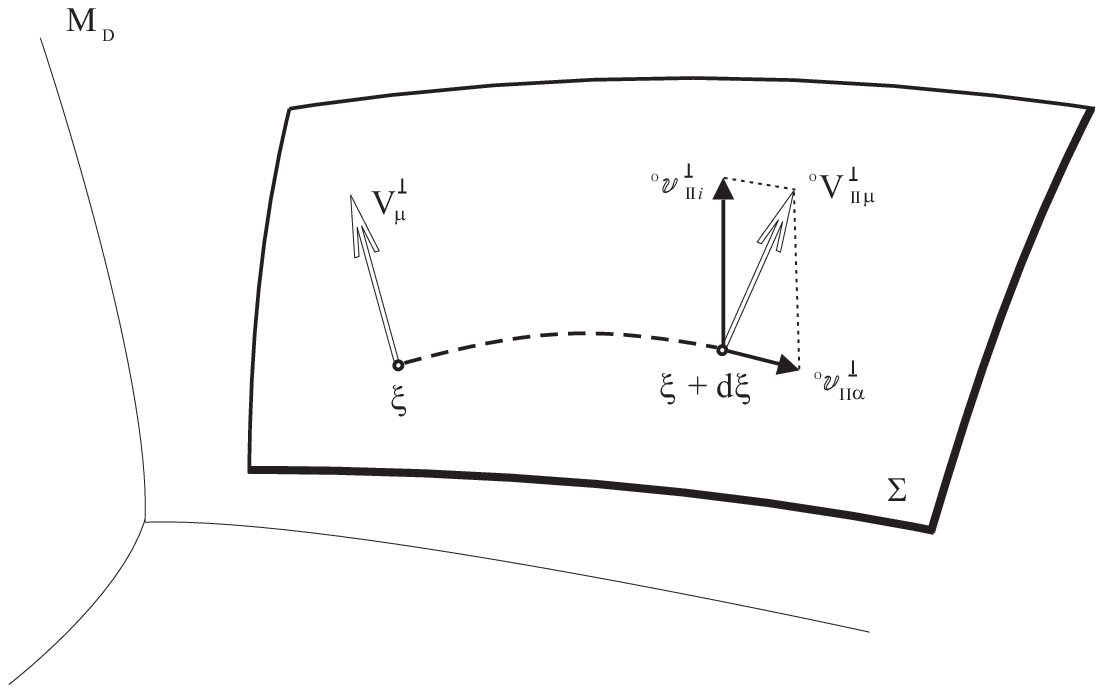,width=13cm}
        \caption[Example]{{\it Definition of the induced connection} ${}^\circ \Gamma^j_{i \alpha}$
        {\it and SFF} ${}^\circ b_{i \alpha \beta}$ {\it from the parallel transport of
        the world-sheet orthogonal covector. Here}
 ${}^\circ v^\bot_{\parallel \alpha} \equiv {}^\circ b_{i \alpha \beta} \, v^i d \xi^\beta
 \,$ {\it and} ${}^\circ  v_{\parallel i}^\bot \equiv v_i + {}^\circ \Gamma^j_{i \alpha}
        v_j \, d \xi^\alpha \,$. }%
    \label{ort}}

Note that the expression for this new SFF
\be
{}^\circ b_{i \alpha \beta} = n_i^\mu  G_{\mu \nu} {}^\circ D_\beta  \partial_\alpha x^\nu
 = -  \partial_\alpha x^\nu  {}^\circ D_\beta ( G_{\mu \nu} n_i^\mu ) \,   ,   \leq{sff}
\ee
differs from the previous one \eq{sff2} by the term containing the covariant
derivative of the metric. So, in the spaces with nonmetricity, there are two
forms of SFF, connected by the relation
\be
{}^\circ b_{i \alpha \beta} = {}^\circ {\bar b}_{i \alpha \beta} +
{}^\circ Q_{\rho \mu \nu} \partial_\beta x^\rho n^\mu_i \partial_\alpha x^\nu  \,   .
\ee

c) {\it Parallel transport of the world-sheet vectors.} In the similar
procedure for the vectors $V^\mu_\Sigma$ and $V^\mu_\bot$, the orthogonal
projection of the ${}^\circ V^\mu_{\Sigma \parallel}$  produce $\, - {}^\circ
b_{i \alpha \beta}$ and the world-sheet projection of ${}^\circ V^\mu_{\bot
\parallel}$ produce ${}^\circ {\bar b}_{i \alpha \beta}$.

In analogy with the previous result, we define the bar induced connection
through the relation
\be
v_\alpha +{}^\circ {\bar \delta} v_\alpha = \partial_\alpha x^\mu (\xi + d \xi)
G_{\mu \nu}(\xi + d \xi) {}^\circ V^\nu _{\Sigma \parallel}  \,  ,  \leq{ptdb}
\ee
where
\be
{}^\circ {\bar \delta} v_\alpha ={}^\circ {\bar \Gamma}^\beta_{\alpha \gamma } v_\beta d \xi^\gamma \,  .  \leq{ptvb}
\ee
Then, we have
\be
{}^\circ {\bar \Gamma}_{\gamma, \alpha \beta} =  \partial_\gamma x^\nu
{}^\circ D_\beta ( G_{\mu \nu} \partial_\alpha x^\mu ) = {}^\circ \Gamma_{\gamma, \alpha \beta}
- {}^\circ Q_{\rho \mu \nu} \partial_\beta x^\rho \partial_\alpha x^\mu \partial_\gamma
x^\nu \,   ,
\ee
which relates the two forms of the induced connections up to the nonmetricity
term.

So, the induced connection and the SFF are projections, of the space-time
covariant derivative of $\partial_\alpha x^\mu $, to the world-sheet and its
normal, respectively. Consequently, we have the generalization of the
Gauss-Weingarten equation
\be
{}^\circ \! D_\beta  \partial_\alpha x^\mu = {}^\circ \Gamma^\gamma_{\alpha \beta}
\partial_\gamma x^\mu  + {}^\circ
b^i_{\alpha \beta} \, n_i^\mu      \,  ,   \leq{scdd}
\ee
and
\be
{}^\circ \! D_\beta ( G_{\mu \nu} \partial_\alpha x^\nu ) =
({}^\circ {\bar \Gamma}^\gamma_{\alpha \beta} \partial_\gamma x^\nu  + {}^\circ
{\bar b}^i_{\alpha \beta} \, n_i^\nu )  G_{\mu \nu}       \,  .   \leq{scddb}
\ee

There are also two forms of the world-sheet induced covariant derivatives,
corresponding to two forms of the world-sheet connections
\be
{}^\circ \nabla_\alpha v_\beta = \partial_\alpha v_\beta - {}^\circ \Gamma^\gamma_{\beta
\alpha} v_\gamma  \, ,  \qquad
{}^\circ {\bar \nabla}_\alpha v_\beta = \partial_\alpha v_\beta - {}^\circ {\bar \Gamma}^\gamma_{\beta
\alpha} v_\gamma  \,   .
\ee
They are related by the expression
\be
{}^\circ {\bar \nabla}_\alpha v_\beta = {}^\circ \nabla_\alpha
v_\beta + {}^\circ Q_{\alpha \beta \gamma} v^\gamma = G_{\beta \gamma} {}^\circ \nabla_\alpha
v^\gamma  \,   .
\ee
From this point we will use the variables without bar, because the
bar variables can be expressed in terms of the first ones and
nonmetricity.

\subsection{Decomposition of the induced connection ${}^\circ \Gamma^\gamma_{\alpha \beta}$
and of the SFF ${}^\circ b_{i \alpha \beta}$ }

In analogy with the general rule for connection decomposition, we
can decompose the induced connection and SFF. With the help of
\eq{cde} we have
\be
{}^\circ D_\beta \partial_\alpha x^\mu = D_\beta \partial_\alpha x^\mu
+ \left( {}^\circ K^\mu{}_{\nu \rho} + \fr{1}{2} {}^\circ Q^{\{ \mu}{}_{\nu \rho
\}} \right)  \partial_\alpha x^\nu  \partial_\beta x^\rho  \,   . \leq{dscd}
\ee
The world-sheet projection of the last equation, produces the
decomposition of the induced connection
\be
{}^\circ \Gamma_{\gamma ,\alpha \beta} = \Gamma_{\gamma ,\alpha \beta} + {}^\circ K_{\gamma \alpha \beta} +
\frac{1}{2} {}^\circ Q_{\{\gamma \alpha \beta \} }   \,   ,     \leq{cdews}
\ee
in terms of induced Christoffel connection, {\bf induced torsion} and
{\bf induced nonmetricity}
\be
\Gamma^\gamma_{\alpha \beta} = G^{\gamma \delta} \partial_\delta x^\mu \, G_{\mu \nu}
{}^\circ \! D_\beta \partial_\alpha x^\nu    \,  , \leq{ich}
\ee
\be
{}^\circ T_{\alpha \beta \gamma} ={}^\circ T_{\mu \rho \sigma} \partial_\alpha x^\mu
\partial_\beta x^\rho \partial_\gamma x^\sigma = {}^\circ \Gamma_{\alpha, \beta \gamma} -
{}^\circ \Gamma_{\alpha, \gamma \beta}   \,  ,   \leq{it}
\ee
\be
{}^\circ Q_{\alpha \beta \gamma} ={}^\circ Q_{\mu \rho \sigma} \partial_\alpha x^\mu \partial_\beta x^\rho
\partial_\gamma x^\sigma = - {}^\circ \nabla_\alpha G_{\beta \gamma}      \,  ,  \leq{inm}
\ee
where ${}^\circ K_{\gamma \alpha \beta} = \fr{1}{2}
{}^\circ T_{\{\gamma \alpha \beta\}}$ is
the induced contortion. Note that both the torsion and
nonmetricity are tensors, so that the corresponding relations do not have
non-homogeneous terms.

The orthogonal projection of \eq{dscd} produces the decomposition
of the SFF
\be
{}^\circ b_{i \alpha \beta} = b_{i \alpha \beta} + {}^\circ \! K_{i \alpha \beta}
+ \fr{1}{2} {}^\circ \! Q_{\{i \alpha \beta \}} \equiv
b_{i \alpha \beta} + {}^\circ \! \stackrel{K}{b}_{i \alpha \beta}
+ \fr{1}{2} {}^\circ \! \stackrel{Q}{b}_{\{i \alpha \beta \}}  \,  ,
\ee
where we used the notation
\be
b_{i \alpha \beta} = n_i^\mu  G_{\mu \nu}  D_\beta  \partial_\alpha x^\nu  \,   ,   \leq{sffCh}
\ee
\be
{}^\circ \! \stackrel{T}{b}_{i \alpha \beta} =
{}^\circ T_{\mu \rho \sigma} n_i^\mu \partial_\alpha x^\rho  \partial_\beta x^\sigma
\equiv  {}^\circ T_{i \alpha \beta}    \,   ,
\ee
\be
{}^\circ \! \stackrel{Q}{b}_{i \alpha \beta} =
{}^\circ \! Q_{\mu \rho \sigma} n_i^\mu \partial_\alpha x^\rho  \partial_\beta x^\sigma
\equiv  {}^\circ \! Q_{i \alpha \beta}    \,   ,
\ee
and similarly as before ${}^\circ K_{i \alpha \beta} = \fr{1}{2}
{}^\circ T_{\{i \alpha \beta\}}$.

Consequently, with the help of \eq{scdd} we have
\be
D_\beta \partial_\alpha x^\mu  =  \Gamma^\gamma_{\alpha \beta}\, \partial_\gamma x^\mu  +
 b^i_{\alpha \beta}\, n_i^\mu   \,   ,
\ee
\be
{}^\circ  T^\mu{}_{\rho \sigma} \partial_\alpha x^\rho \partial_\beta x^\sigma
\equiv  {}^\circ T^\mu{}_{\alpha \beta} =
{}^\circ T^\gamma{}_{\alpha \beta} \partial_\gamma x^\mu  +
{}^\circ \! T^i_{\alpha \beta}\, n_i^\mu   \,   ,
\ee
\be
{}^\circ \! Q^\mu {}_{\rho \sigma} \partial_\alpha x^\rho  \partial_\beta x^\sigma
\equiv  {}^\circ \! Q^\mu{}_{\alpha \beta} =
{}^\circ Q^\gamma{}_{\alpha \beta} \partial_\gamma x^\mu  +
{}^\circ \! Q^i_{\alpha \beta}\, n_i^\mu   \,  .
\ee

When the torsion is present, the SFF is not symmetric in
$\alpha,\beta$ indices. Similarly as in the case of the
connection, we can write
\be
{}^\circ b_{i \alpha \beta} = {}^\circ b_{i ( \alpha \beta)} +
\fr{1}{2} {}^\circ T_{i \alpha \beta} \,  ,  \qquad
{}^\circ b_{i (\alpha \beta)} = b_{i \alpha \beta} + {}^\circ K_{i (\alpha
\beta)} + \fr{1}{2} {}^\circ Q_{\{ i \alpha \beta \} } \,   .
\ee

\subsection{Relations between space-time and world-sheet covariant derivatives}

Starting with the definition of the space-time covariant derivatives
along world-sheet direction
\be
{}^\circ D V^\mu = V^\mu ( \xi + d \xi) - V^\mu_{\parallel} \,   ,
 \leq{dcd}
\ee
we can obtain its world-sheet and orthogonal projections,
multiplying \eq{dcd} with $G_{\mu \nu} (\xi + d \xi) \partial_\alpha x^\nu ( \xi + d \xi)$
and $G_{\mu \nu} (\xi + d \xi) n^\nu_i ( \xi + d \xi)$, respectively
\be
\partial_\alpha x^\nu G_{\mu \nu} {}^\circ \! D_\beta V^\mu =
{}^\circ {\bar \nabla}_\beta v_\alpha - v^i \, {}^\circ \! {\bar b}_{i \alpha
\beta}  \,  ,   \qquad
n_i^\nu G_{\mu \nu} {}^\circ D_\alpha V^\mu =
{}^\circ {\bar \nabla}_\alpha v_i + v^\beta \, {}^\circ b_{i \beta \alpha}   \,  .
\ee
With the help of the relation $V^\mu = \partial_\alpha x^\mu
v^\alpha+ n_i^\mu v^i$ we obtain
\be
{}^\circ D_\alpha V^\mu = {}^\circ \nabla_\alpha v^\beta \, \partial_\beta x^\mu +
n_i^\mu \, {}^\circ b^i_{\beta \alpha} v^\beta + n_i^\mu G^{i j} \, {}^\circ {\bar
\nabla}_\alpha v_j - \partial_\beta x^\mu v^i G^{\beta \gamma}\, {}^\circ \! {\bar b}_{i \gamma
\alpha} \,   .  \leq{rcdu}
\ee
Similarly, for the covectors we have
\be
{}^\circ D_\alpha V_\mu = G_{\mu \nu} \left[ G^{\beta \gamma} \partial_\beta x^\nu {}^\circ \nabla_\alpha v_\gamma +
n_i^\nu \, {}^\circ {\bar b}^i_{\beta \alpha} v^\beta \right]
+ G_{\mu \nu} \left[G^{i j} n_i^\nu  {}^\circ \nabla_\alpha v_j -
G^{\beta \gamma} \partial_\beta x^\nu v^i \, {}^\circ b_{i \gamma
\alpha} \right] \,   .   \leq{rcdd}
\ee

For the world-sheet tangent vectors, the last two equations
produce
\be
{}^\circ D_\alpha V^\mu_\Sigma = {}^\circ \nabla_\alpha v^\beta \, \partial_\beta x^\mu +
n_i^\mu \, {}^\circ b^i_{\beta \alpha} v^\beta  \,  ,   \leq{Ge}
\ee
\be
{}^\circ D_\alpha V_\mu^\Sigma = G_{\mu \nu} \left[ G^{\beta \gamma}
\partial_\beta x^\nu \, {}^\circ \nabla_\alpha v_\gamma +
n_i^\nu \, {}^\circ {\bar b}^i_{\beta \alpha} v^\beta \right]  \,   .  \leq{Ged}
\ee
We will also need the relation
\be
({}^\circ D_\alpha V_\mu) \partial_\beta x^\mu = {}^\circ \nabla_\alpha
v_\beta - v^i \, {}^\circ b_{i \beta \alpha} \,  ,  \leq{rcd}
\ee
obtaining multiplying \eq{rcdd} by $\partial_\beta x^\mu$.

\subsection{Mean extrinsic curvature}

In the torsion free case, the SFF is symmetric in the world-sheet indices and
its properties are well known in the literature. Here, we will consider the
case when nontrivial torsion and nonmetricity are present.

As usual, a curve is parametrized by its length parameter $s$, so that the
unit tangent vector is $t^\mu = {d x^\mu \over ds}$. If the curve lies in the
world-sheet, we have $t^\mu =  t^\alpha \partial_\alpha x^\mu$, with $t^\alpha
= {d \xi^\alpha \over ds}$. Let us denote by $P_i$, the 2-plane spanned by the
tangent vector $t^\mu$ and the unit world-sheet normal  $n_i^\mu$. Then, $\ell_i
= \Sigma \bigcap P_i$ is the $i$-th normal section of the world-sheet
$\Sigma$. The curvature of the normal section $\ell_i$, as a space-time curve,
is defined as the orthogonal projection of ${}^\circ D_s t^\mu$
\be
{}^\circ k_i = {}^\circ D_s t^\mu G_{\mu \nu} n^\nu_i  \,  ,
\ee
where ${}^\circ D_s t^\mu \equiv {\dot x}^\nu {}^\circ D_\nu t^\mu$, is  the
covariant derivative along the curve. It produces the following expression
\be
{}^\circ k_i = {}^\circ  b_{i \beta \alpha}\, t^\alpha t^\beta =
{}^\circ  b_{i (\beta \alpha)}\, t^\alpha t^\beta = { {}^\circ
b_{i (\beta \alpha)} d \xi^\alpha d \xi^\beta \over G_{\alpha
\beta} d \xi^\alpha d \xi^\beta} \,  .       \leq{cns}
\ee
The curvature ${}^\circ k_i$ depends on the direction of the tangent vector $
t^\alpha$ and only on the symmetric part of the SFF. Let us stress, that in
the presence of nonmetricity, ${}^\circ D_s t^\mu$ is not orthogonal to the
tangent vector $t^\mu$, and we additionally performed orthogonal projection.

In Euclidean torsion free spaces, the maximal and minimal values of ${}^\circ
\! k_i$ are the principal curvatures. The corresponding directions, defined by
$t^\alpha$, are the principal directions. The principal curvatures are
eigenvalues of the SFF and corresponding eigenvectors define the principal
directions.

In our case, the curvature ${}^\circ \! k_i$,  \eq{cns}, is divergent in the
light-cone directions. Consequently, the extremely values do not exist. Still,
we can obtain the necessary information from the eigenvalue problem
\be
({}^\circ \! b^i_{\alpha \beta} -  {}^\circ \! \kappa^i G_{\alpha
\beta}) v^\beta = 0  \,   .           \leq{ssp}
\ee

The eigenvalues of the quadratic forms ${}^\circ  b^i_{\alpha \beta}$ with
respect to the metric $G_{\alpha \beta}$, we define as a principal
curvatures in Minkowski space-time. They are the solutions of the condition
$\det({}^\circ \! b^i - {}^\circ \!\kappa^i G)_{\alpha \beta} = 0$, or explicitly
\be
{}^\circ \! \kappa^i_{0,1} =  {}^\circ \! H^i \pm \sqrt{ ({}^\circ \! H^i)^2 -  {}^\circ \!
K^i} \,    .
\ee
Here
\be
{}^\circ \! H^i = \fr{1}{2} G^{\alpha \beta} {}^\circ \! b^i_{\alpha \beta} \,
 =  \fr{1}{2} ({}^\circ \! \kappa^i_0 + {}^\circ
\! \kappa^i_1)  \,   ,  \leq{aec}
\ee
is the trace of the SFF and ${}^\circ \! K^i = { \det {}^\circ \! b^i_{\alpha \beta} \over
\det G_{\alpha \beta}}= {}^\circ \! \kappa^i_0 {}^\circ \! \kappa^i_1 $ (no
summation over $i$) is Gauss curvature.

In order to simplify the relation \eq{cns}, we will go to the new frame in
which both symmetric matrices, $G_{\alpha \beta}$ and ${}^\circ b_{i (\alpha
\beta)}$, are diagonal. We can choose the eigenvectors
($v_0^\alpha$, $v_1^\alpha$) of the symmetric part of the SFF ${}^\circ \!
b^i_{(\alpha \beta)}$, as new basis vectors. Then, instead of \eq{ssp} we have
$({}^\circ \! b^i_{(\alpha \beta)} - {}^\circ \! \kappa^i G_{\alpha \beta})
v^\beta = 0$, which produces $({}^\circ \! \kappa^i_0 - {}^\circ \! \kappa^i_1
)(v_0^\alpha G_{\alpha \beta} v_1^\beta)=0$. For ${}^\circ \! \kappa^i_0 \neq
{}^\circ \! \kappa^i_1$, the eigenvectors are orthogonal. Let us chose $v_0$
to be time-like vector and $v_1$ to be space-like vector, and normalize them
as $v_0^2 =1 ,\, \, v_1^2 = -1$. In the basis ($v_0, v_1$), both the metric
and the symmetric part of the SFF, obtain diagonal forms
\be
 G = \pmatrix{1  & 0 \cr
           0    & -1     \cr }  \,  , \qquad
b_i = \pmatrix{{}^\circ \! \kappa^i_0    & 0 \cr
           0    &  - {}^\circ \! \kappa^i_1      \cr }  \,   .
\ee
The curvature of the line $\ell_i$ becomes
\be
{}^\circ k^i_{tl} (\alpha) = {}^\circ \! \kappa^i_0 \cosh^2(\alpha )
- {}^\circ \! \kappa^i_1 \sinh^2 (\alpha)  \,   ,
\ee
where $\alpha \in (- \infty, \infty)$ parametrize all time-like directions of
the tangent vectors. Similarly, for the space-like tangent vectors, we have
\be
{}^\circ k^i_{sl} (\alpha) = {}^\circ \! \kappa^i_0 \sinh^2(\alpha )
- {}^\circ \! \kappa^i_1 \cosh^2 (\alpha)  \,   .
\ee

The $i$-th MEC usually is defined as a mean value of the $i$-th normal
section, if we fixed the normal $n_i^\mu$ and rotate the tangent vector
$t^\mu$. In Euclidean spaces, the $\sin$ and $\cos$ functions appear instead
of $\sinh$ and $\cosh$, so that the mean value is well defined. In Minkowski
spaces, the curvature \eq{cns} is divergent in the light-like directions. We
introduce the cut-off $\Lambda$, and define regularized mean curvature for the
time-like region
\be
{}^\circ H^i_{tl} (\Lambda) = {1 \over 2 \Lambda} \int_{-
\Lambda}^\Lambda {}^\circ k^i_{tl} (\alpha) d \alpha \,   ,
\ee
and similarly for the space-like one. Finally, the {\bf mean
extrinsic curvature} we define as a mean value of the time-like
mean curvature with sign $+$, and space-like mean curvature with sign
$-$, with the same cut-off $\Lambda$ going to the infinity.
We obtain
\be
 \lim_{\Lambda \to \infty} \fr{1}{2} [{}^\circ H^i_{tl}
(\Lambda)- {}^\circ H^i_{sl} (\Lambda)] =
\fr{1}{2} ({}^\circ \! \kappa^i_0 + {}^\circ
\! \kappa^i_1) = {}^\circ \! H^i    \,   ,  \leq{aec1}
\ee
which is just the expression \eq{aec}, because the $\Lambda$ dependence
disappears.

For ${}^\circ \! \kappa^i_0 = {}^\circ \! \kappa^i_1 \equiv {}^\circ \!
\kappa^i$, all linear combinations of the vectors $v_0$ and $v_1$ are
eigenvectors. We can choose some orthonormal basis, in which
\be
G = \pmatrix{1  & 0 \cr
           0    & -1     \cr }  \,  , \qquad
b_i = \pmatrix{{}^\circ \! \kappa^i    & 0 \cr
           0    &  - {}^\circ \! \kappa^i     \cr }  \,   .
\ee
Then the line curvature does not depend on the tangent vector direction
\begin{eqnarray}
{}^\circ k^i_{tl} (\alpha)& = &{}^\circ \! \kappa^i [ \cosh^2(\alpha )
-  \sinh^2 (\alpha)] = {}^\circ \! \kappa^i = {}^\circ \! H^i_{tl}
\,  ,    \nonumber  \\
{}^\circ k^i_{sl} (\alpha) &= &{}^\circ \! \kappa^i [\sinh^2(\alpha )
-  \cosh^2 (\alpha)] = -{}^\circ \! \kappa^i = {}^\circ \! H^i_{sl}   \,   .
\end{eqnarray}
Consequently, the MEC has the same interpretation
$\fr{1}{2} ({}^\circ \! H^i_{tl} - {}^\circ \!
H^i_{sl}) = {}^\circ \! \kappa^i = {}^\circ \! H^i$.

The surface defined by the equation ${}^\circ \! H^i = 0$ is
{\bf minimal surface}. The name stems from the fact that in the Riemann
space-time the equation $H_i = 0$ define the surface of minimal area
$P_2 = \int d^2 \xi \sqrt{-\det G_{\alpha \beta}}\,$, for the fixed boundary. In fact, for this
surface the first variation of the area vanishes
\be
\delta P_2 = -2  \int d^2 \xi \sqrt{-\det G_{\alpha \beta}} \, H^i n^\mu_i G_{\mu \nu} \, \delta x^\nu
= 0  \,  .
\ee

\subsection{Dual mean extrinsic curvature, extrinsic mean torsion and C-duality}

Let us stress that above consideration is torsion independent, because the
antisymmetric part of the SFF disappears from  \eq{cns} and  \eq{aec}. We are
going to include the torsion contribution and generalize the above results.

Let us first generalize the eigenvalue problem, and then offer its geometrical
interpretation. We introduce the {\bf dual eigenvalue } problem, such that
linear transformation of the vector $v^\alpha$ with operator ${}^\circ \!
b^i_{\alpha \beta}$ is proportional to the two dimensional dual vector
${}^\ast v_\alpha = \sqrt{-G_2} \, \varepsilon_{\alpha \beta} \, v^\beta \,\,
(G_2 = \det G_{\alpha \beta})$
\be
{}^\circ \! b^i_{\alpha \beta} \, v^\beta = {}^\ast \kappa^i \, {}^\ast v_\alpha
\,    ,
\ee
or equivalently
\be
\left( {}^\circ \! b^i_{\alpha \beta} -  {}^\ast \kappa^i \, \varepsilon_{\alpha
\beta} \sqrt{-G_2} \right) v^\beta = 0  \,   .           \leq{asp}
\ee
It is similar to \eq{ssp}, but for the completeness we also need the
eigenvalues of the quadratic forms ${}^\circ  b^i_{\alpha
\beta}$ with respect to the antisymmetric tensor $\varepsilon_{\alpha \beta}
\sqrt{-G_2}$.

The solutions of the condition $\det( {}^\ast \! b^i -  {}^\ast k^i \,
\varepsilon \sqrt{-G_2})_{\alpha \beta} = 0$, have the form
\be
{}^\ast \kappa^i_{0,1} =  {}^\ast H^i \pm \sqrt{({}^\ast H^i)^2 + {}^\circ \!
K^i} \,    .
\ee
In analogy with the previous case, we will call them dual principal curvatures,
and the variable
\be
{}^\ast H^i = \fr{1}{2} ({}^\ast \kappa^i_0 + {}^\ast \kappa^i_1) =
\fr{1}{2} G^{\alpha \beta} {}^\ast  b^i_{\alpha \beta} =
\fr{1}{2} {\varepsilon^{\alpha \beta} \over \sqrt{-G_2}} {}^\circ \! b^i_{\beta \alpha} =
\fr{1}{4} {\varepsilon^{\alpha \beta} \over \sqrt{-G_2}}
{}^\circ \! T^i_{\beta \alpha}     \,  ,          \leq{dmc}
\ee
the {\bf dual mean extrinsic curvature}. Here ${}^\circ \! K^i =
{ \det {}^\circ \! b^i_{\alpha \beta} \over \det G_{\alpha
\beta}}= {}^\ast \kappa^i_0 {}^\ast \kappa^i_1 $ (no summation over $i$) is the
same Gauss curvature as before.

Let us now turn to the geometrical meaning of the DMEC. In the case when
$t^\mu_1$ and $t^\mu_2$ are world-sheet tangent vectors (Fif. 1) (note that
all geodesics $\ell_1$, $\ell_2$, $\tilde{\ell}_1$ and $\tilde{\ell}_2$ still
could be space-time curves) we can rewrite  \eq{cod} in the form
\be
{}^\circ T^\mu  \equiv  {x^\mu(D_2) - x^\mu(D_1) \over 2 d P_{12}} =
{\varepsilon^{\beta \alpha} \over 4 \sqrt{-G_2}}\,
{}^\circ T^\mu_{\rho \sigma} \partial_\alpha x^\rho  \partial_\beta x^\sigma  \,    .
\ee
Here, $ d P_{12} = \sqrt{-G_2} \det( {\partial
\xi^\alpha \over \partial s_r}) d \ell_1 d \ell_2$ is area of the
parallelogram, spanned by the vectors $\ell_1^\mu = t_1^\mu d
\ell_1$ and $\ell_2^\mu = t_2^\mu d \ell_2\,$, and where $G_2 = \det G_{\alpha
\beta}$. We can conclude that ${}^\circ T^\mu$
does not depend on the directions $t^\mu_1$ and $t^\mu_2$, and on the
lengths $d \ell_1$ and $d \ell_2$. So, we will call this variable the {\bf mean torsion}.
Its world-sheet projection is induced mean torsion
\be
{}^\circ T_\gamma  = {}^\circ T^\mu G_{\mu \nu} \partial_\gamma x^\nu =
 {\varepsilon^{\beta \alpha} \over 4 \sqrt{-G_2}}\, {}^\circ T_{\gamma \alpha \beta}   \,   .
\ee
Its normal projection is the {\bf extrinsic mean torsion}
\be
{}^\circ T_i  = {}^\circ T^\mu G_{\mu \nu} n_i^\mu =
{\varepsilon^{\beta \alpha} \over 4 \sqrt{-G_2}} \, {}^\circ T_{i \alpha \beta}
=  {}^\ast H_i   \,  ,
\ee
which is exactly the same variable as DMEC, defined above in \eq{dmc}.

We can formulate the dual eigenvalue problem   \eq{asp}, as an ordinary
eigenvalue problem $({}^\ast b^i_{\alpha \beta}  - {}^\ast \kappa^i G_{\alpha
\beta} ) v^\alpha = 0$, if we introduce the dual SFF
\be
{}^\ast b^i_{\alpha \beta} = {G_{\alpha \gamma}
\varepsilon^{\gamma\delta} \over  \sqrt{-G_2} } \, {}^\circ \! b^i_{\delta
\beta} \,  .
\ee

We define the {\bf C-duality} ({\bf C}urvature duality), which maps SFF to
dual SFF, ${}^\circ b^i_{\alpha \beta} \to {}^\ast b^i_{\alpha \beta}$, and
interchanges the role played by the symmetric and the antisymmetric parts of
the SFF. Consequently, under C-duality MEC maps to DMEC, allowing the exchange
of the mean curvature and mean torsion.

The self-dual and self-antidual configurations
\be
{}^\circ H^i = \pm  {}^\ast H^i  \,      ,      \leq{sdr}
\ee
correspond to the following conditions on the SFF
\be
(G^{\alpha \beta} \mp {\varepsilon^{\alpha \beta} \over
\sqrt{-G_2}}) {}^\circ \! b^i_{\beta \alpha} = 0  \,  .   \leq{sasd}
\ee
The equations  \eq{sdr} and \eq{sasd} define {\bf C-dual (antidual) surfaces}.
In the torsion free case, they turn to the standard minimal surface condition,
${}^\circ H^i= 0$.

\section{The stringy geometry as a space-time perception by probe string}

We are going to investigate the dynamics of the string, propagated in the
curved space-time. From the mathematical point of view, the  string equations of
motion describe the embedding conditions of the world-sheet in the space-time.
We are particularly interested in the target space geometry properties,
recognized by the string. We will see, that the probe string feels more
space-time features (torsion and nonmetricity) then the probe particle.

\subsection{The action and the canonical analysis results}

The action \cite{GSW}-\cite{BNS}
\be
S= \kappa  \int_\Sigma d^2 \xi \sqrt{-g} \left\{ \left[ {1 \over 2}g^{\alpha\beta}G_{\mu\nu}(x)
+{\varepsilon^{\alpha\beta} \over \sqrt{-g}}  B_{\mu\nu}(x)\right]
\partial_\alpha x^\mu \partial_\beta x^\nu + \Phi(x) \stackrel{\con}{R}^{(2)} \right\}  \, ,   \leq{ac}
\ee
describes bosonic string propagation in $x^\mu$-dependent background fields:
metric $G_{\mu \nu}$, antisymmetric tensor field $B_{\mu\nu}=- B_{\nu\mu}$ and
dilaton field $\Phi$. Here, $g_{\alpha \beta}$ is the intrinsic world-sheet
metric and $\stackrel{\con}{R}^{(2)}$ is corresponding scalar curvature. In
App. A, we will introduce the intrinsic connection $\con$, as a Christoffel
connection for the intrinsic metric $g_{\alpha \beta}$. So, we mark all
related variables with the sign $\con$, in order to distinguish them from the
corresponding induced ones.

In this paper we will restrict our consideration to such forms of the
dilaton fields $\Phi$, that its gradient $a_\mu = \partial_\mu
\Phi$, is not light-cone vector. So, the condition $a^2 \equiv G^{\mu \nu} a_\mu a_\nu  \neq 0$
is fulfilled in the whole space-time.

The projection operator
\be
P^T{}_{\mu \nu} = G_{\mu \nu} - {a_\mu a_\nu \over a^2} = G_{\mu \nu}- \varepsilon n_\mu n_\nu \equiv G_{\mu \nu}^{D-1} \,  ,   \leq{PT}
\ee
($n_\mu = {a_\mu \over \sqrt{\varepsilon a^2}}$, where $\varepsilon=1$ if
$a_\mu$ is time like vector and $\varepsilon=-1$ if $a_\mu$ is space like
vector), can be interpreted as the induced metric on the $D-1$ dimensional
submanifold $\Phi(x)= const$.

Let us briefly review the result of the canonical analysis of ref. \cite{BS}.
It is useful to define the currents
\be
J_{\pm \mu} =P^T{}_\mu{}^\nu j_{\pm \nu} +{a_\mu \over 2
a^2}i_\pm^\Phi = j_{\pm \mu} -{ a_\mu \over a^2} j \,  ,  \leq{J}
\ee
\be
i_\pm^F= {a^\mu \over a^2} j_{\pm \mu}-{1 \over 2 a^2} i_\pm^\Phi
\pm 2 \kappa {F^\prime } \,  , \qquad i_\pm^\Phi= \pi_F \pm 2\kappa
\Phi' \,  ,   \leq{2.8}
\ee
where
\be
j_{\pm \mu} =\pi_\mu +2\kappa \Pi_{\pm \mu \nu} {x^\nu}' \,  ,
\qquad  \Pi_{\pm \mu \nu} \equiv  B_{\mu \nu} \pm {1 \over 2}
G_{\mu \nu} \,  , \leq{jmi} \ee and
\be
j=a^\mu j_{\pm \mu} -{1 \over 2} i_\pm^\Phi =a^2 (i_\pm^F \mp 2 \kappa F^\prime)  \,   .   \leq{j}
\ee

All $\tau$ and $\sigma$ derivatives of the fields $x^\mu$, $F$ and $\Phi$ can be
expressed in terms of the corresponding currents
\be
{\dot x}^\mu =  {G^{\mu \nu} \over 2 \kappa} (h^- J_{- \nu} - h^+ J_{+ \nu}) \,  ,  \leq{xt}
\ee
\be
{\dot F}= {1 \over 4 \kappa} (h^- i_-^F - h^+ i_+^F)-{1 \over 2} { (h^-+h^+)^\prime} \, ,  \qquad
{\dot \Phi}= {1 \over 4 \kappa} (h^- i_-^\Phi - h^+ i_+^\Phi)  \,  ,       \leq{Ft}
\ee
and
\be
x^{\mu \prime}= {G^{\mu \nu} \over 2 \kappa} (J_{+ \nu} - J_{- \nu})  \, , \qquad
{F^\prime }= {1 \over 4 \kappa} (i_+^F - i_-^F)   \,  ,  \qquad
{\Phi^\prime }= {1 \over 4 \kappa} (i_+^\Phi - i_-^\Phi)  \,   .   \leq{prim}
\ee

Up to boundary term, the canonical Hamiltonian density has the standard form
\be
{\cal H}_c= h^- T_- + h^+ T_+    \,   .  \leq{hc}
\ee
The energy momentum tensor components obtain new expressions
\be
T_\pm =\mp {1 \over 4\kappa} \left(G^{\mu\nu} J_{\pm \mu} J_{\pm \nu} + i_\pm^F i_\pm^\Phi \right)
+{1 \over 2} i_\pm^{\Phi \prime} =
\mp {1 \over 4\kappa} \left( G^{\mu\nu}
j_{\pm \mu} j_{\pm \nu} -{j^2 \over a^2} \right) +i_\pm  \,   ,  \leq{emt}
\ee
where
\be
i_\pm= {1 \over 2} ({i_\pm^\Phi}' - F' i_\pm^{\Phi}) \,  .  \leq{2.16}
\ee

The same chirality energy-momentum tensor components satisfy two independent
copies of Virasoro algebras,
\be
\{ T_\pm , T_\pm \}= -[ T_\pm(\sigma) +T_\pm({\bar \sigma}) ] { \delta^\prime} \,   ,   \leq{Vir}
\ee
while the opposite chirality components commute $\{ T_\pm , T_\mp \}= 0$. The
energy-momentum tensor components $T_\pm$ are generators of the two
dimensional diffeomorphisms. Their new, non-linear representation is the
consequence of the dilaton field presence.

\subsection{Equations of motion}

In the paper \cite{BS}, using canonical approach, we derived the following
equations of motion
\be
[J^\mu]  \equiv  \stackrel{\con}{\nabla}_\mp \partial_\pm  x^\mu + {}^\star \Gamma_{\mp \rho \sigma}^\mu
\partial_\pm x^\rho \partial_\mp x^\sigma =0  \,  ,    \leq{lJ}
\ee
\be
[h^\pm]  \equiv  G_{\mu \nu}  \partial_\pm  x^\mu \partial_\pm x^\nu -2 \stackrel{\con}{\nabla}_\pm \partial_\pm \Phi =0   \,  ,  \leq{lh}
\ee
\be
[i^F]   \equiv  \stackrel{\con}{R}^{(2)} + {2 \over a^2} (D_{\mp \mu} a_\nu) \partial_\pm  x^\nu \partial_\mp  x^\mu =0   \,  ,     \leq{lF}
\ee
where the variables in the parenthesis denote the currents corresponding
to this equation. The expression
\be
{}^\star \Gamma^\rho_{\pm \nu \mu}= \Gamma^\rho_{\pm \nu \mu} +{a^\rho \over a^2} D_{\pm \mu} a_\nu
= P^{T \rho}{}_\sigma \Gamma^\sigma_{\pm \nu \mu} +
{a^\rho \over a^2} \partial_\mu a_\nu = \Gamma^\rho_{\nu \mu} \pm P^{T \rho}{}_\sigma B^\sigma_{\nu \mu}
+{a^\rho \over a^2} D_{\mu} a_\nu   \,  ,  \quad  \leq{cdc}
\ee
which appears in the $[J^\mu]$ equation is a generalized connection, which
full geometrical interpretation will be presented later. Under space-time
general coordinate transformations the expression ${}^\star \Gamma^\rho_{\pm
\nu \mu}$ transforms as a connection. In  \eq{lJ} and \eq{lF} we omit the
currents $\pm$ indices, because $[J^\mu_+] = [J^\mu_-]$ and $[i^F_+] =
[i^F_-]$ as a consequence of the symmetry relations ${}^\star \Gamma_{\mp \rho
\sigma}^\mu = {}^\star \Gamma_{\pm \sigma \rho}^\mu$ and $D_{\mp \mu} a_\nu =
D_{\pm \nu} a_\mu$.

We can derive the Lagrangian equations of motion, independently of the previous
consideration. Varying the action \eq{ac} with respect to $x^\mu$, we obtain
\be
[x^\mu]  \equiv g^{\alpha \beta} \stackrel{\con}{\nabla}_\alpha \partial_\beta  x^\mu + \Gamma^{(\alpha \beta) \mu}{}_{\rho \sigma}
\partial_\alpha x^\rho \partial_\beta x^\sigma  -a^\mu \stackrel{\con}{R}^{(2)} = 0  \,
,   \leq{emx}
\ee
where
\be
\Gamma^{(\alpha \beta) \mu}{}_{\rho \sigma} = g^{\alpha \beta} \,  \Gamma^\mu_{\rho \sigma} -
{\varepsilon^{\alpha \beta} \over \sqrt{-g}} B^\mu_{\rho \sigma}  \,  .  \leq{fie}
\ee

Because in the light-cone basis we have $g^{\pm \pm} =0$ and $\varepsilon^{\pm \pm} =0$, the above expression has
only two nonzero components
\be
\Gamma^{(\mp \pm ) \mu}{}_{\rho \sigma} =  \Gamma^\mu_{\pm \rho \sigma}   \, ,  \qquad
 \Gamma^{(\pm \pm ) \mu}{}_{\rho \sigma} = 0  \, .  \leq{lx}
\ee
So, the expression "two types of connections", which has been used in the literature, means
that in the light-cone basis there are just two nonzero elements of five
indices expression \eq{fie}. The connections $\Gamma^\mu_{\pm \rho \sigma}$
corresponds to the parallel transport along the light-cone lines
$d \xi^\pm$, respectively. Trough the paper we will preserve the word
connection, having in mind this comment.

Instead of the equations of motion with respect to the world-sheet metric
$g_{\alpha \beta}$, we prefer to have ones with respect to the fields $F$ and
$h^\pm$
\be
[F] \equiv \stackrel{\con}{\Delta} \Phi =0 \, ,   \leq{emF}
\ee
\be
T_\pm = \mp \frac{\kappa}{2} (G_{\mu \nu}  {\partial}_\pm x^\mu {\partial}_\pm x^\nu
-2 \stackrel{\con}{\nabla}_\pm {\partial}_\pm \Phi - \stackrel{\con}{ \Delta} \Phi )=0  \,  ,  \leq{lemh}
\ee
where the $T_\pm$ are energy-momentum tensor components in the light-cone basis.

Let us compare these equations with ones obtained before, using canonical
methods. The equation  \eq{lh} follows from \eq{emF} and \eq{lemh}. The
equation  \eq{emx} in the light-cone basic obtains the form
\be
[x^\mu_\pm] \equiv   \stackrel{\con}{\nabla}_\mp {\partial}_\pm  x^\mu +  \Gamma_{\mp \rho \sigma}^\mu  {\partial}_\pm x^\rho
{\partial}_\mp x^\sigma -  \fr{1}{2} a^\mu \stackrel{\con}{R}^{(2)}  =0   \,  ,
\ee
because $[x^\mu] = [x^\mu_+] + [x^\mu_-]$ and $[x^\mu_+] = [x^\mu_-]$ as a
consequence of the above symmetry relations. Then we can conclude that $[i^F]$
and $[J^\mu]$ contain longitudinal and transversal parts of $[x^\mu]$
respectively
\be
[i^F] = - {1 \over a^2} \left( a_\mu [x^\mu] + [F] \right)   \,  , \qquad 2[J^\mu] = P^{T
\mu}{}_\nu [x^\nu] + {a^\mu \over a^2} [F]    \,  .
\ee

We prefer the equations of motion in the form \eq{lJ}-\eq{lF}, with the
canonical origin, because they are more
appropriate for geometrical interpretation.

\subsection{Stringy torsion and nonmetricity}

Let us apply the general considerations of Sec. 2 to the string case. Instead
of general mark ${}^\circ$, the mark ${}^\star$ will indicates the presence of
nonmetricity felt by the string, and lower indices  $\pm$ will indicate the
presence of the corresponding form of torsion felt by the string. Absence of
any sign will means that the connection is Christoffel.

The manifold $M_D$, together with the affine connection ${}^\circ
\Gamma^\mu_{\nu \rho}$ and the metric $G_{\mu \nu}$, define the affine
space-time $A_D \equiv ( M_D, {}^\circ \Gamma, G)$. The connection \eq{cdc} we
will call {\bf stringy connection} and the corresponding space-time $S_D
\equiv (M_D, {}^\star \Gamma_\pm, G)$, observed by the string propagating in
the background $G_{\mu \nu}$, $B_{\mu \nu}$ and $\Phi$, we will call {\bf
stringy space-time}. The classification of space-times dependence on the
background fields contributions, will be investigated in Sec. 4.

The antisymmetric part of the stringy connection is the {\bf stringy torsion}
\be
{}^\star T_\pm{}^\rho_{\mu \nu} ={}^\star \Gamma_\pm{}^\rho_{\mu \nu} - {}^\star \Gamma_\pm{}^\rho_{\nu \mu} =
\pm 2 P^{T \rho}{}_\sigma  B^\sigma_{\mu \nu}      \,  .   \leq{T}
\ee
It is the transverse projection of the field strength of the antisymmetric
tensor field $B_{\mu \nu}$.

The presence of the dilaton field $\Phi$ leads to the breaking of the
space-time metric postulate. It means that the metric $G_{\mu \nu}$ is not
compatible with the stringy connection ${}^\star \Gamma^\mu_{\pm \nu \rho}$.
This fact can be expressed by the nontrivial {\bf stringy nonmetricity}
\be
{}^\star Q_{\pm \mu \rho \sigma} \equiv  -{}^\star D_{\pm \mu} G_{\rho \sigma} =
{1 \over a^2} D_{\pm \mu} (a_\rho a_\sigma )       \,   .   \leq{mp}
\ee
Consequently, during  stringy parallel transport, the lengths and angles
deformations depend on the vector field $a_\mu$.

Note that ${}^\star \Gamma^\rho_{\pm \nu \mu}$, ${}^\star T^\rho_{\pm \nu \mu}$ and ${}^\star Q_{\pm \mu \rho \sigma}$
are invariant under scale transformation of the dilaton field  by the constant $ \Phi \to k  \Phi$.

The stringy Weyl vector
\be
{}^\star Q_\mu  = {1 \over D} G^{\rho \sigma } {}^\star Q_{\pm \mu \rho \sigma}= {-4 \over D} \partial_\mu \varphi    \,    ,     \leq{Wvs}
\ee
is a gradient of the new scalar field $\varphi$, defined by the expression
\be
\varphi=-{1 \over 4} \ln a^2 = -{1 \over 4} \ln(G^{\mu \nu} \partial_\mu \Phi
\partial_\nu \Phi) \,  .   \leq{sf}
\ee
It does not depend on the antisymmetric field $B_{\mu \nu}$ and consequently on the $\pm$ indices.

The stringy angle preservation relation
\be
{}^\star \sQ_{\pm \mu \rho \sigma} = {}^\star Q_{\pm \mu \rho \sigma} - G_{\rho \sigma} {}^\star Q_\mu =0     \,       ,   \leq{aps}
\ee
is a condition on the dilaton field $\Phi$. Generally, in stringy geometry both
the lengths and the angles could be changed under the parallel transport.

Using the relation
\be
{}^\star K_{\pm \mu \rho \sigma} + \frac{1}{2} {}^\star Q_{\pm \{ \mu \rho \sigma \} } =
\pm \frac{1}{2} {}^\star T_{\mu \rho \sigma} + \frac{1}{2} {}^\star Q_{\{ \mu \rho \sigma \} }  \,   ,
\ee
instead \eq{cde}, we can write
\be
{}^\star \Gamma_{\pm \mu ,\rho \sigma} = \Gamma_{\mu ,\rho \sigma} \pm \frac{1}{2} {}^\star T_{\mu \rho \sigma} +
\frac{1}{2} {}^\star Q_{\{ \mu \rho \sigma \} }   \,   ,     \leq{cde1}
\ee
where the quantities ${}^\star T_{\mu \rho \sigma} = 2  P^T{}_\mu^\nu B_{\nu \rho \sigma}$ and
${}^\star Q_{ \mu \rho \sigma} = -{}^\star D_{\mu} G_{\rho \sigma} = {1 \over a^2} D_{\mu} (a_\rho
a_\sigma )$ do not depend on $\pm$ indices. In fact, the last term is ${}^\star Q_{\{ \mu \rho
\sigma \} } = 2 \frac{a_\mu}{a^2} D_\rho a_\sigma$, so that we can recognize our starting
expression \eq{cdc}.

In the stringy geometry, covectors $a_\mu$ are covariantly constant target space vectors
\be
{}^\star D_{\pm \mu} a_\nu =0 \,    .     \leq{cva}
\ee
By definition it means that ${}^\star \delta a_\mu = d a_\mu$, or that the
covector $a_\mu(x)$, after parallel transport from the point $x$ to the point
$x+dx$ with the connection ${}^\star \Gamma^\mu_{\pm \rho \sigma}$, is equal
to the local covector $a_\mu(x+dx)$. So, covector field $a_\mu$ is {\it
stringy teleparallel}, because its parallel transport is path independent.
Note that as a consequence of nonmetricity, the above feature valid only for
covectors $a_\mu$, and not for the vectors $a^\mu$.

\section{ Background fields contribution to the space-time geometry }

In this section we will make the classification of space-time, depending on
the presence of the background fields. We will analyze the field equations and
find their geometrical interpretations.

\subsection{Riemann space-time induced by metric $G_{\mu \nu}$}

Let us start with the case where only nontrivial background field is the
metric tensor $G_{\mu \nu}$, while $B_{\mu \nu}=0=\Phi $. Then, instead of
$J_{\pm \mu}$, the current takes the form $ j_{\pm \mu}^G =\pi_\mu \pm \kappa
G_{\mu \nu}  x^{\nu \prime}$, but with the same Lagrangian expression
$j^G_{\pm \mu}= \sqrt{2} \kappa G_{\mu \nu} {\hat \partial}_\pm x^\nu$. The
canonical Hamiltonian has the similar form as in \eq{hc}
\be
{\cal H}_c^G= h^- t_-^G + h^+ t_+^G   \,   ,  \leq{2.141}
\ee
with following expression for the energy-momentum tensor
\be
t^G_\pm =\mp {1 \over 4\kappa} G^{\mu\nu} j_{\pm \mu}^G j_{\pm \nu}^G \,  .   \leq{emtG}
\ee

The absence of the dilaton field $\Phi$, leads to the conformal invariance of
the action. Consequently, the field $F$ and the corresponding equation
$[i^F]$ are absent. The first two equations of motion \eq{lJ} and \eq{lh}
survive in the simpler form
\be
[j_G^\mu]  \equiv \, \stackrel{\con}{\nabla}_\mp \partial_\pm x^\mu + \Gamma_{\rho \sigma}^\mu
\partial_\pm x^\rho  \partial_\mp x^\sigma =0    \, ,  \qquad
[h^\pm]  \equiv G_{\mu \nu}  \partial_\pm  x^\mu \partial_\pm x^\nu =0 \,   ,
\ee
where $\Gamma^\mu_{\rho \sigma}$ is the standard Christoffel
connection.

The world-sheet projection and orthogonal projection of the equation
$[j^\mu_G]$,  obtain the forms
\be
g^{\alpha \beta} (\con^\gamma_{ \alpha \beta } - \Gamma^\gamma_{ \alpha \beta })=0  \,   ,  \leq{ic}
\ee
and
\be
g^{\alpha \beta} b^i_{\beta \alpha} = 0  \,  .   \leq{imc}
\ee
Here, $\Gamma_{\alpha \beta}^\gamma$ is induced world-sheet Christoffel
connection \eq{ich} and $b^i_{\alpha \beta}$ is corresponding SFF \eq{sffCh}.

The $[h^\pm]$ equation can be written in the form
\be
G_{\pm \pm} \equiv e_\pm^\alpha e_\pm^\beta  G_{\alpha \beta} = 0     \,  , \leq{imtem}
\ee
where $G_{\alpha \beta} = G_{\mu \nu} \partial_\alpha x^\mu \partial_\beta x^\nu $ is world-sheet
induced metric. Because in the light-cone frame, the intrinsic metric is off-diagonal,
$g_{\pm \pm} =0$, we have $G_{a b} = \lambda g_{a b}, \,\, (a,b \in \{+, -\})$, or equivalently
\be
G_{\alpha \beta } = \lambda g_{\alpha \beta}     \,      .  \leq{icc}
\ee
Multiplying the last equation with $g^{\alpha \beta}$, we obtain the
expression $ \lambda = \fr{1}{2}g^{\alpha \beta}G_{\alpha \beta }$, so that
$\sqrt{-G} = \sqrt{-g} \fr{1}{2}g^{\alpha \beta}G_{\alpha \beta }$. The last
relation connects the Polyakov and Nambu-Goto expressions for the string
action. We can rewrite \eq{icc} in the form
\be
{G_{\alpha \beta } \over \sqrt{-G}} =  {g_{\alpha \beta } \over \sqrt{-g}} \,  ,
\ee
so that, because of the conformal invariance, only the metric densities are related.

From \eq{icc} we obtain relation between intrinsic connection
$\con^\gamma_{ \alpha \beta }$ and induced one $\Gamma^\gamma_{ \alpha \beta }$
\be
\con^\gamma_{ \alpha \beta } = \Gamma_{\alpha \beta}^\gamma -\delta^\gamma_\alpha \lambda_\beta -
\delta^\gamma_\beta \lambda_\alpha + g_{\alpha \beta} \, g^{\gamma \delta} \lambda_\delta \,  ,
\qquad (\lambda_\alpha \equiv \fr{1}{2} \partial_\alpha \ln \lambda)   \leq{cc}
\ee
which is also a solution of \eq{ic}. Therefore, both intrinsic world-sheet
metric tensor and intrinsic connection, are equal to the induced ones from the
space-time up to the conformal factor $\lambda$. This is expected result,
because of the conformal invariance of the action. The complete equality of
the metric tensors and connections is just the choice of the gauge fixing
$\lambda=1$.

With the help of  \eq{icc} equation \eq{imc} becomes
\be
H^i \equiv \fr{1}{2}G^{\alpha \beta} b^i_{\beta \alpha} = 0 \,  .
\ee
Therefore, all MECs, corresponding to the normal vectors $n_i^\mu$, are equal
to zero. In the geometrical language the  world-sheet $\Sigma$ is {\it minimal
surface}. From starting $D$ equations of motion, two define intrinsic
connection in terms of the induced one and $D-2$ define the MEC.

The string with the background field $G_{\mu \nu}$ does not  see the torsion
and nonmetricity. It feels the target space as {\it Riemann space-time} of
general relativity \cite{MB,HMMN}.

\subsection{Stringy Riemann-Cartan space-time induced by metric
$G_{\mu \nu}$ and antisymmetric tensor  $B_{\mu \nu}$}

In the next step, we include antisymmetric background field $B_{\mu \nu}$, but
still keep  $\Phi=0$. Now, the relevant current $j_{\pm \mu}$ is already
defined by expression \eq{jmi}, with the same form of Hamiltonian  \eq{hc} and
with the energy-momentum tensor
\be
t_\pm =\mp {1 \over 4\kappa} G^{\mu\nu} j_{\pm \mu} j_{\pm \nu}  \,  .   \leq{emt0}
\ee
With the same reason as in the previous subsection, we have again two equations of motion
\be
[j^\mu]  \equiv \, \stackrel{\con}{\nabla}_\mp \partial_\pm x^\mu + \Gamma_{\mp \rho \sigma}^\mu
\partial_\pm x^\rho \partial_\mp x^\sigma =0  \,     , \qquad
[h^\pm] \equiv  G_{\mu \nu}  \partial_\pm  x^\mu \partial_\pm x^\nu =0  \,  .  \leq{emjh}
\ee
Now, two forms of connection
\be
\Gamma^\rho_{\pm \nu \mu}= \Gamma^\rho_{\nu \mu} \pm B^\rho_{\nu \mu}    \,  ,  \leq{gco}
\ee
and corresponding two forms of covariant derivative $D_{\pm \mu}$, appear. The
new term in the connection is the field strength of the antisymmetric tensor,
defined in \eq{fsB}. Note that $G_{\mu \nu}$ and $B_{\mu \nu}$ always appear
in the combination $\Pi_{\pm \mu \nu}= B_{\mu \nu} \pm \frac{1}{2}G_{\mu
\nu}$, so that the generalized connection \eq{gco} can be expressed as in
\eq{coP}.

The torsion
\be
T_\pm{}^\rho_{\mu \nu} = \Gamma_\pm{}^\rho_{\mu \nu} - \Gamma_\pm{}^\rho_{\nu \mu} =
\pm 2 B^{\rho}_{\mu \nu} \,  ,           \leq{Tor}
\ee
is the field strength for the antisymmetric tensor. In this case the
contortion is proportional to the torsion $K_{\pm \mu \nu \rho} = \fr{1}{2}
T_{\pm \mu \nu \rho} = \pm B_{\mu \nu \rho} $.

As well as in the case of Riemann space-time, the same relations between
metric tensors \eq{icc} and between the connections \eq{cc}, follow from the
$[h^\pm]$ equation  \eq{emjh}.

We can rewrite the $[j^\mu]$ equation in the form
\be
g^{\alpha \beta} ( D_\beta \partial_\alpha x^\mu - \con^\gamma_{\alpha
\beta} \partial_\gamma x^\mu) = { \varepsilon^{\alpha \beta} \over
\sqrt{-g}}B^\mu_{\alpha \beta}  \,  .
\ee
Its world-sheet projection produce
\be
g^{\alpha \beta} (\con^\gamma_{\alpha \beta} - \Gamma^\gamma_{ \alpha \beta})
= - { \varepsilon^{\alpha \beta} \over \sqrt{-g}}B^\gamma_{\alpha
\beta}    \,  ,   \leq{cei}
\ee
where $B^\gamma_{\alpha \beta}$ is the world-sheet torsion, induced from the target
space. Because it is totally antisymmetric, in two dimensions it vanishes,
$B^\gamma_{\alpha \beta} =0$. So, we obtain the same equation  \eq{ic}, as
in the case of Riemann space-time.

The orthogonal projection of the $[j^\mu]$ equation takes the form
\be
g^{\alpha \beta} b_{i \alpha \beta} =  { \varepsilon^{\alpha \beta} \over
\sqrt{-g}} B_{i \alpha \beta}  \, ,   \leq{dem}
\ee
or equivalently
\be
(g^{\alpha \beta} \pm { \varepsilon^{\alpha \beta} \over \sqrt{-g}})
b_{\pm i \beta \alpha} = 0   \, .   \leq{dem1}
\ee
We used two forms of SFF $b_{\pm i \alpha \beta}=  n_i^\mu G_{\mu
\nu} D_{\pm \beta} \partial_\alpha  x^\nu$, which are connected
$b_{\pm i \alpha \beta} = b_{\mp i \beta \alpha}$. So, \eq{dem1} contains
only one independent equation of motion.

Again, both two dimensional intrinsic metric tensor and two dimensional
intrinsic connection, up to the conformal factor, are induced from the target
space. So, we can rewrite the above equation in terms of induced metric
\be
( G^{\alpha \beta} \pm { \varepsilon^{\alpha \beta} \over
\sqrt{-G_2}}) b_{\pm i \beta \alpha} = 0 \,  ,  \qquad
\Longleftrightarrow    \qquad   H_i \pm  {}^\ast H_{\pm i} =0     \,      . \leq{sdrc}
\ee
There is unique MEC $H_i$ and two types of DMECs ${}^\ast H_{\pm i}$,
satisfying the relation ${}^\ast H_{\pm i} = - {}^\ast H_{\mp i}$. So, the
equations \eq{sdrc} with upper and lower indices are equal. In this language
equation of motion means that MEC is equal to DMEC, or that world-sheet
$\Sigma$ is {\it C-dual (antidual) surface}.

Consequently, the string with background fields $G_{\mu \nu}$ and $B_{\mu
\nu}$ feels the target space as {\it Riemann-Cartan  space-time}
\cite{MB,HMMN}. Note that the world-sheet is torsion free while the space-time
is not.

\subsection{Stringy torsion free space-time induced by background fields $G_{\mu \nu}$ and $\Phi$}

Let us consider the case when the fields $G_{\mu \nu}$ and $\Phi$ are present,
 while $B_{\mu \nu}$ is absent. The equations of motion \eq{lJ}-\eq{lF} obtain
the form
\be
[J^\mu]  \equiv \, \stackrel{\con}{\nabla}_\mp \partial_\pm  x^\mu + {}^\star \Gamma_{\rho \sigma}^\mu
\partial_\pm x^\rho \partial_\mp x^\sigma =0  \,  ,    \leq{lJgf}
\ee
\be
[h^\pm]  \equiv  G_{\mu \nu}  \partial_\pm  x^\mu \partial_\pm x^\nu -2 \stackrel{\con}{\nabla}_\pm \partial_\pm \Phi =0   \,  ,  \leq{lhgf}
\ee
\be
[i^F]  \equiv \, \stackrel{\con}{R}^{(2)} + {2 \over a^2} (D_\mu a_\nu) \partial_\mp x^\mu \partial_\pm x^\nu = 0   \,  ,     \leq{lFgf}
\ee
and can be rewritten as
\be
[J^\mu]  \equiv \, g^{\alpha \beta} \left( \stackrel{\con}{\nabla}_\beta \partial_\alpha  x^\mu +
{}^\star \Gamma_{\rho \sigma}^\mu  \partial_\alpha x^\rho \partial_\beta x^\sigma \right) =0  \,  ,    \leq{lJgf1}
\ee
\be
[h^\pm]  \equiv G_{\pm \pm} - 2 \stackrel{\con}{\nabla}_\pm a_\pm = 0   \,  ,  \leq{lhgf1}
\ee
\be
[i^F]  \equiv \, \stackrel{\con}{R}^{(2)} + {g^{\alpha \beta} \over a^2} (D_\alpha a_\mu) \partial_\beta x^\mu  = 0   \,  .     \leq{lFgf1}
\ee

The connection
\be
{}^\star \Gamma^\rho_{ \nu \mu}= \Gamma^\rho_{\nu \mu} + {a^\rho \over a^2} D_\mu a_\nu \,  ,  \leq{ctf}
\ee
obtains the new term, which is the origin of the nonmetricity
\be
{}^\star Q_{ \mu \rho \sigma} \equiv  -{}^\star D_\mu G_{\rho \sigma} =
{1 \over a^2} D_\mu (a_\rho a_\sigma )       \,   ,   \leq{mptf}
\ee
while the torsion is zero.

Let us find the induced connection and the SFF, for this case. With the help
of \eq{ctf} we have
\be
{}^\star D_\beta \partial_\alpha x^\mu  = D_\beta \partial_\alpha x^\mu
+ {a^\mu \over a^2} (D_\beta a_\nu) \partial_\alpha x^\nu   \,   .
\ee
The world-sheet projection produce the induced connection
\be
{}^\star \Gamma^\gamma_{\alpha \beta} = \Gamma^\gamma_{\alpha
\beta} + {a^\gamma \over a^2} (D_\beta a_\mu) \partial_\alpha x^\mu   \,   ,   \leq{stfco}
\ee
and the normal projection produce the SFF
\be
{}^\star b^i_{\alpha \beta} = b^i_{\alpha \beta} +
{a^i \over a^2} (D_\beta a_\mu) \partial_\alpha x^\mu     \,   .
\ee

Now, we are going to derive the useful relation between the space-time and the
world-sheet covariant derivatives, which we will frequently use later.

Substituting mark ${}^\circ$, with mark ${}^\star$ in \eq{rcd} and separating
$a_\mu$ dependent terms, we obtain
\be
\nabla_\alpha v_\beta = ( D_\alpha V_\mu - { a^\rho V_\rho^\perp \over a^2} D_\alpha a_\mu )
\partial_\beta x^\mu  + v^i \, {}^\star b_{i \beta \alpha} \,   .  \leq{cv}
\ee
Here, $V_\mu^\perp$ is the component of the covector $V_\mu$ orthogonal to the
world-sheet. For $V_\mu \to a_\mu$ we have
\be
{1 \over a^2} (D_\alpha a_\mu) \partial_\beta x^\mu  = {1 \over a_2^2}
\nabla_\alpha a_\beta - {a^i \over a_2^2} {}^\star b_{i \beta \alpha}  \,  ,   \leq{pcd}
\ee
where $a_2^2 = G^{\alpha \beta} a_\alpha a_\beta$ is a length of the world-sheet projection of the
vector field $a_\mu$. We also used the relation $a^2=a_2^2 +aa^\perp$.

With the help of  \eq{pcd}, the stringy torsion free connection
\eq{stfco} becomes
\be
{}^\star  \Gamma^\gamma_{\alpha \beta} = \Gamma^\gamma_{\alpha
\beta} + {a^\gamma \over a_2^2} \nabla_\beta a_\alpha  -  {a^\gamma a^i \over
a_2^2} {}^\star b_{i \alpha \beta}  \,  . \leq{sic}
\ee
The stringy torsion free world-sheet covariant derivative is
\be
{}^\star \nabla_\alpha v_\beta = \nabla_\alpha v_\beta - {(av)_2
\over a_2^2} (\nabla_\alpha a_\beta - a^i {}^\star b_{i \beta \alpha})  \,  ,   \leq{scd}
\ee
and for $v_\beta \to a_\beta$, it produces
\be
{}^\star \nabla_\alpha a_\beta = a^i \, {}^\star b_{i \beta \alpha}   \,   . \leq{scda}
\ee

Let us turn to the equations of motion. The world-sheet and orthogonal
projections of $[J^\mu]$ equation are
\be
g^{\alpha \beta} ({}^\star  \Gamma^\gamma_{\alpha \beta} -
\con^\gamma_{\alpha \beta}) =0 \  ,   \qquad  \Longleftrightarrow    \qquad
g^{\alpha \beta} ( \stackrel{\con}{\nabla}_\alpha v_\beta -
{}^\star \nabla_\alpha v_\beta)= 0  \,   ,  \leq{fsJ}
\ee
\be
g^{\alpha \beta} \, {}^\star b_{i \alpha \beta } = 0  \,   .   \leq{ssJ}
\ee
The last equation, together with \eq{scda} produce $g^{\alpha
\beta} {}^\star \nabla_\alpha a_\beta = 0$, and with the help of
second equation \eq{fsJ} we obtain $\stackrel{\con}{\nabla}_\pm a_\mp = 0$.

We will preserve the condition $G_{\pm \pm}=0$, or equivalently $G_{\alpha
\beta} = \lambda g_{\alpha \beta}$, because the light-cone line must be the
same in terms of the intrinsic and induced metrics. Then, the new term in
$[h^\pm]$ equation vanishes $\stackrel{\con}{\nabla}_\pm a_\pm = 0$, which
complete the above condition
\be
\stackrel{\con}{\nabla}_\alpha a_\beta = 0  \,   .  \leq{vca}
\ee

As well as in the previous two cases, from the relation $G_{\alpha \beta} =
\lambda g_{\alpha \beta}$ follows the equalities of the intrinsic and induced
metric densities. The relation between corresponding connections changes in
the presence of the nonmetricity
\be
{}^\star Q_{\alpha \beta \gamma} = ({}^\star \Gamma - \lambda \con)_{\beta, \gamma \alpha}
+ ({}^\star \Gamma - \lambda \con)_{\gamma, \beta \alpha} -
2 \lambda_\alpha G_{\beta \gamma}  \, .  \qquad (\lambda_\alpha \equiv
\fr{1}{2} \partial_\alpha \ln \lambda)  \,     \leq{inml}
\ee
From \eq{sic} and \eq{vca} we have $a^\beta ({}^\star \Gamma - \lambda \con)_{\beta, \gamma
\alpha}= - a^i {}^\star b_{i \gamma \alpha} $ and consequently
\be
{}^\star Q_{\alpha \beta \gamma} a^\beta a^\gamma = -2 a^i a^\beta {}^\star b_{i \beta
\alpha} - 2 a_2^2 \lambda_\alpha\,  .  \leq{inm1}
\ee

On the other hand, from \eq{mptf}, \eq{inm} and \eq{pcd} we obtain
\be
{}^\star Q_{\alpha \beta \gamma} = {1 \over a_2^2} \nabla_\alpha(
a_\beta a_\gamma) - {a^i \over a_2^2}( a_\beta {}^\star b_{i \gamma \alpha} +
 a_\gamma {}^\star b_{i \beta \alpha})    \,  ,  \leq{wsinm}
\ee
and
\be
{}^\star Q_{\alpha \beta \gamma} a^\beta a^\gamma =
\partial_\alpha a_2^2 -2 a^i a^\beta {}^\star b_{i \beta \alpha}       \,  .  \leq{inm2}
\ee
Subtracting \eq{inm1} from \eq{inm2},  we obtain the equation for $\lambda
\,$, $\, \partial_\alpha \ln \lambda = - \partial_\alpha \ln a_2^2$. Its
solution is easily to be found, $\lambda = { 1 \over a_2^2}$. Therefore, when
the dilaton field $\Phi$ is present, the theory lose conformal invariance and
the coefficient $\lambda$ is determined. In our case, we have relation between
intrinsic and induced metric tensors
\be
g_{\alpha \beta} = a_2^2 G_{\alpha \beta}   \,  ,   \leq{gg}
\ee
and between intrinsic and induced connections
\be
\con^\gamma_{\alpha \beta} = {}^\star \Gamma^\gamma_{\alpha \beta} - \delta_\alpha^\gamma
\lambda_\beta  - \delta_\beta^\gamma \lambda_\alpha + G_{\alpha \beta} G^{\gamma \delta}
\lambda_\delta - \fr{1}{2} {}^\star Q^{\{ \gamma}{}_{\alpha \beta
 \}} \,  .    \qquad   \left( \lambda_\alpha \equiv -\partial_\alpha \ln
 \sqrt{a_2^2} \right)
\ee

With the help of \eq{pcd} and \eq{ssJ}, from the equation
$[i^F]$ we obtain the expression for the intrinsic world-sheet curvature
\be
\stackrel{\con}{R}^{(2)} = - {G^{\alpha \beta} \over a_2^4}
\nabla_\alpha a_\beta   \,    .  \leq{iwsc}
\ee
Let us find its relation with the induced one. Both curvatures we can
relate with the hat curvature ${\hat R}$
\be
\sqrt{-g}  \stackrel{\con}{R}^{(2)} = \sqrt{- {\hat g}}( {\hat R}
+ 2 {\hat \Delta} F) \,  , \qquad
\sqrt{-G_2}  R^{(2)} = \sqrt{- {\hat g}}( {\hat R}
+ 2 {\hat \Delta} F_G ) \,  ,
\ee
where in agreement with  \eq{g} we use light-cone variables for
both intrinsic and induced metrics $g_{\alpha \beta} \to (h^\pm,
F), \, G_{\alpha \beta} \to (h^\pm, F_G)$. Eliminating ${\hat R}$
and using \eq{gg} (which is equivalent to the relation
$F_G = F - \fr{1}{2} \ln a_2^2$) we have
\be
R^{(2)}=  a_2^2  \stackrel{\con}{R}^{(2)} - \Delta \ln  a_2^2    \,  , \leq{tdcr}
\ee
where $\Delta$ is Laplace operator for the induced metric
$\Delta = e^{-2 F_G} {\hat \Delta}$. So, in our case
with the help of  \eq{iwsc} we have
\be
R^{(2)}=  {1 \over a_2^4} \Delta \Phi - \Delta \ln  a_2^2   \,   .   \leq{itdc}
\ee

The equation \eq{ssJ} with the help of \eq{gg} takes the form
\be
{}^\star H_i  = \fr{1}{2} G^{\alpha \beta} \, {}^\star b_{i \beta \alpha} = 0  \,
.    \leq{stfe}
\ee
Consequently, all stringy MECs, ${}^\star H_i$, vanish and the world-sheet is
{\it stringy minimal surface}.

Therefore, the string with background fields $G_{\mu \nu}$ and $\Phi$ can see
the space-time nonmetricity. The corresponding target space we call {\it
stringy torsion free space-time}.

\subsection{Stringy space-time induced by background fields $G_{\mu \nu}$, $B_{\mu \nu}$ and $\Phi$}

Finally, in the last step we include the dilaton field $\Phi$, so that all
three background fields are present.

The equations of motion are the complete ones, \eq{lJ}-\eq{lF}, with two forms
of connections \eq{cdc}. The string feel both torsion and nonmetricity defined
by the expressions \eq{T} and \eq{mp} respectively. Separating the $a_\mu$
dependence, we can rewrite the field equations in the form
\be
[J^\mu]  \equiv \, g^{\alpha \beta} \left({}^\star D_\beta \partial_\alpha  x^\mu
- \con^\gamma_{\alpha \beta} \partial_\gamma x^\mu \right) -
{\varepsilon^{\alpha \beta} \over \sqrt{-g}} P^{T \mu} {}_\nu B^\nu
{}_{ \alpha \beta} = 0     \,  ,    \leq{lJs}
\ee
\be
[h^\pm]  \equiv G_{\pm \pm} - 2 \stackrel{\con}{\nabla}_\pm a_\pm = 0   \,  ,  \leq{lhs}
\ee
\be
[i^F]  \equiv \, \stackrel{\con}{R}^{(2)} + {g^{\alpha \beta} \over a^2} (D_\alpha a_\mu) \partial_\beta x^\mu
+ {\varepsilon^{\alpha \beta} \over \sqrt{-g} \, a^2} a_\mu  B^\mu {}_{ \alpha \beta} = 0      \,  .     \leq{lFs}
\ee

We will follow the considerations of the previous two cases. First, we find
the consequences of the relation \eq{rcd} substituting mark ${}^\circ$ with
marks ${}^\star {}_\pm$ and separating the $a_\mu$ dependent terms. The
contribution of the $B^{T \mu} {}_{\alpha \beta}$ term disappears and we
obtain the same relations \eq{cv} and \eq{pcd}.

The stringy connection and stringy world-sheet covariant derivatives are
\be
{}^\star \Gamma^\gamma_{\pm \alpha \beta} = {}^\star \Gamma^\gamma_{\alpha \beta}
\pm B^{T \gamma}{}_{\alpha \beta}  \,   , \qquad
{}^\star \nabla_{\pm \alpha} v_\beta = {}^\star \nabla_{\alpha} v_\beta
\mp  B^{T \gamma}{}_{\beta \alpha} v_\gamma  \,   ,
\ee
where ${}^\star \Gamma^\gamma_{\alpha \beta}$ and ${}^\star \nabla_{\alpha}
v_\beta$ are defined in \eq{sic} and \eq{scd}, so that \eq{scda}
still valid because $B^{T \gamma}{}_{\beta \alpha}$ disappears.

The stringy world-sheet induced nonmetricity has a form
\be
{}^\star Q_{\pm \alpha \beta \gamma} = {}^\star Q_{\alpha \beta \gamma}\mp
(B^L_{\beta \alpha \gamma} + B^L_{\alpha \beta \gamma})    \,    ,   \leq{wsinmpm}
\ee
where ${}^\star Q_{\alpha \beta \gamma}$ has been introduced in \eq{wsinm},
and $B^L_{\beta \alpha \gamma} = \partial_\beta x^\nu {a_\nu a_\mu \over a^2}
B^\mu {}_{\alpha \gamma}$ is a world-sheet projection of the longitudinal part
of the field strength $ B^\mu {}_{\alpha \gamma}$.

The world-sheet and the normal projections of the $[J^\mu]$ equation are
\be
g^{\alpha \beta} ({}^\star  \Gamma^\gamma_{\alpha \beta} -
\con^\gamma_{\alpha \beta}) = {\varepsilon^{\alpha \beta} \over \sqrt{-g}}
B^{T \gamma} {}_{ \alpha \beta} = \mp 2 {\sqrt{-G_2} \over
\sqrt{-g}} \, \, {}^\star T^\gamma_\pm     \  ,       \leq{fsBJ}
\ee
\be
g^{\alpha \beta} \, {}^\star b_{i \alpha \beta } =  {\varepsilon^{\alpha \beta} \over \sqrt{-g}}
B^T {}_{i \alpha \beta}  \, ,  \qquad \Longleftrightarrow \qquad
\left( g^{\alpha \beta}  \pm {\varepsilon^{\alpha \beta} \over
\sqrt{-g}} \right) {}^\star b_{\pm i \beta \alpha} = 0    \,   .   \leq{ssBJ}
\ee
Using \eq{scda}, \eq{fsBJ} and \eq{ssBJ} we obtain
\be
g^{\alpha \beta} \stackrel{\con}{\nabla}_\alpha a_\beta =
{\varepsilon^{\alpha \beta} \over \sqrt{-g}} (a^i B^T {}_{i \alpha
\beta}+ a^\gamma B^T {}_{\gamma \alpha \beta}) = {\varepsilon^{\alpha \beta} \over
\sqrt{-g}} \, a^\mu B^T {}_{\mu \alpha \beta} = 0    \,   .  \leq{vca1}
\ee
This is the same result as in the torsion free case,
$\stackrel{\con}{\nabla}_\pm a_\mp = 0$.

For the same reason as in the previous case, we will preserve the condition
$G_{\pm \pm}=0$, which is equivalent to the relation $G_{\alpha \beta} =
\lambda g_{\alpha \beta}$. Together with the $[h^\pm]$ equation it gives
$\stackrel{\con}{\nabla}_\pm a_\pm = 0$, and finally the same equation
\eq{vca}, $\stackrel{\con}{\nabla}_\alpha a_\beta =0$.

The relation between the connections is similar to \eq{inml}
\be
{}^\star Q_{\pm \alpha \beta \gamma} = ({}^\star \Gamma_\pm - \lambda \con)_{\beta, \gamma \alpha}
+ ({}^\star \Gamma_\pm - \lambda \con)_{\gamma, \beta \alpha} -
2 \lambda_\alpha G_{\beta \gamma}  \, ,  \qquad (\lambda_\alpha \equiv
\fr{1}{2} \partial_\alpha \ln \lambda)  \,     \leq{inmls}
\ee
but now we have $a^\beta ({}^\star \Gamma_\pm - \lambda \con)_{\beta, \gamma
\alpha}= - a^i {}^\star b_{i \gamma \alpha} \pm a^\beta B^T_{\beta \gamma
\alpha}$, so that
\be
{}^\star Q_{\pm \alpha \beta \gamma} a^\beta a^\gamma = -2 a^i a^\beta {}^\star b_{i \beta
\alpha} - 2 a_2^2 \lambda_\alpha  \pm 2 B^T_{\beta \gamma \alpha}  a^\beta a^\gamma  \,  .  \leq{inms1}
\ee
The equation \eq{wsinmpm} produce
\be
{}^\star Q_{\pm \alpha \beta \gamma} a^\beta a^\gamma =
\partial_\alpha a_2^2 -2 a^i a^\beta {}^\star b_{i \beta \alpha} \mp
2 B^L_{\beta \gamma \alpha}  a^\beta a^\gamma      \,  .  \leq{inms2}
\ee
Since
\be
B^T_{\beta \gamma \alpha} + B^L_{\beta \gamma \alpha} = B_{\beta \gamma
\alpha}= 0  \,   ,   \leq{TL}
\ee
because it is totally antisymmetric two-dimensional tensor, we have the same
equation for $\lambda$ with the same solution \eq{gg}.

With the help of \eq{pcd} and \eq{ssBJ} the $[i^F]$ equation becomes
\be
\stackrel{\con}{R}^{(2)} = - {G^{\alpha \beta} \over a_2^4} \nabla_\alpha a_\beta
+ {\varepsilon^{\alpha \beta} \over \sqrt{-g}}
\left( {a^i \over a_2^2} B^T_{i \alpha \beta} - {a^\mu \over a^2} B^T_{\mu \alpha
\beta} \right)  \,  .    \leq{iwscs}
\ee
Using the properties $ a^i B^T_{i \alpha \beta} + a^\gamma  B^T_{\gamma \alpha
\beta}= a^\mu B^T_{\mu \alpha \beta}=0$ and \eq{TL}, we again have
\eq{iwsc}. Consequently, from \eq{tdcr} we obtain \eq{itdc}, as well as in the
torsion free case.

The equation of motion \eq{ssBJ} with the help of \eq{gg} takes the form
\be
\left( G^{\alpha \beta} \pm { \varepsilon^{\alpha \beta} \over
\sqrt{-G_2}} \right) {}^\star b_{\pm i \beta \alpha} = 0   \,  ,
\qquad  \Longleftrightarrow   \qquad    {}^\star H_i \pm  {}^\ast H_{ \pm i} = 0    \,  .         \leq{ssde}
\ee
So, the world-sheet $\Sigma$ is {\it stringy C-dual (antidual) surface}. For
${}^\star Q_{\mu \rho \sigma} = 0$ it turns to  \eq{sdrc}  and for ${}^\star
T_{\mu \rho \sigma} = 0$ to \eq{stfe}.

The string propagating in the presence of all three background fields $G_{\mu
\nu}$, $B_{\mu \nu}$ and $\Phi$, feels both space-time torsion and
nonmetricity. The corresponding target space we call {\it stringy space-time}.

There are some other possibilities which we will not consider here. For
example, the presence of $B_{\mu \nu}$ field only, will lead to the flat
space-time with torsion, which is known as {\it teleparallel} space-time.

\section{The space-time measure}

Using the stringy geometry considered in the previous sections, let us try to
discuss possible forms of the space-time actions. Generally it has the form
\be
{}^\star S = \int d^D x \,\, {}^\star \Omega \,\, {}^\star {\cal L}  \,   ,  \leq{sta}
\ee
where ${}^\star \Omega$ is a measure factor, and ${}^\star {\cal L}$ is a
Lagrangian, which depends on the space-time field strengths.

We are going to find invariant measure, which means that:

1. it is invariant under space-time general coordinate transformations;

2. it is preserved under parallel transport;

3. it enable integration by parts.

The second requirement is equivalent to the condition ${}^\star D_{\pm \mu} {}^\star \Omega = 0$.
The third one, is consequence of the Leibniz rule, and the relation
\be
\int d^D x \, {}^\star \Omega \, {}^\star D_{\pm \mu} V^\mu = \int d^D x \partial_\mu ({}^\star \Omega V^\mu)  \,   ,  \leq{ibp}
\ee
so that we are able to use Stoke's theorem.

For Riemann and Riemann-Cartan space-times, the solution for the measure factor
is well known $\Omega = \sqrt{-G}$ $(G = \det G_{\mu \nu})$. For spaces with
nonmetricity, this standard measure is not preserved under the stringy
parallel transport, and requirements 2. and 3. are not satisfy. Instead to
change the connection and find volume-preserving one, as has been done in ref.
\cite{HMMN}, we prefer to change the measure. Let us try to find the the
stingy measure in the form ${}^\star \Omega= \Lambda(x) \sqrt{-G}$. In order
to be preserved under the parallel transport with the stingy connection, it must
satisfy the condition
\be
{}^\star D_{\pm \mu} ( \sqrt{-G} \Lambda ) = \partial_\mu ( \sqrt{-G} \Lambda ) -
{}^\star \Gamma^\rho_{\pm \mu \rho} \sqrt{-G} \Lambda  = 0       \,  .
\ee
Using the relation
\be
{}^\star \Gamma^\rho_{\pm \mu \rho} = \partial_\mu \ln \left( \sqrt{- G }e^{-2 \varphi}\right) =
\Gamma^\rho_{\pm \mu \rho} + \frac{D}{2} {}^\star Q_\mu         \,  ,  \leq{Grr}
\ee
we find the equation for $\Lambda \,$,  $\,\,\partial_\mu \Lambda =
\frac{D}{2} {}^\star Q_\mu \Lambda$. The fact that the stringy Weyl vector
${}^\star Q_\mu$ is a gradient of the scalar field $\varphi$, defined in
\eq{sf}, help us to solve this equation obtaining $\Lambda = e^{-2 \varphi}$.
The stringy measure factor, preserved under parallel transport
with the connection ${}^\star \Gamma^\mu_{\pm \nu}$, obtains the form
\be
{}^\star \Omega = \sqrt{-G} e^{-2 \varphi}    \,      .
\ee

Note that now we have ${}^\star \Gamma^\rho_{\pm \mu \rho} =  \partial_\mu \ln
{}^\star \Omega$, and consequently \eq{ibp} is satisfied. So, we can use the
integration by parts for stringy derivatives ${}^\star D_{\pm \mu}$, if we use
the stringy measure ${}^\star \Omega$. Therefore, all requirements are
satisfy.

Let us shortly compare the action \eq{sta} with the space-time action of the
papers  \cite{CFMP,BNS}. In spite of their different origin there are some
interesting similarities. The Lagrangians from these papers are scheme
dependent (see \cite{Ts}), and can be reproduced with suitable combinations of
our invariants: the stringy scalar curvature, stringy torsion and stringy
nonmetricity.

We are particularly interested to compare the integration measures. The
measure factor of the papers \cite{CFMP,BNS}, $\sqrt{-G} e^{-2 \Phi}$, has the
same form as our one and confirm the existence of some space-time
nonmetricity. The requirement of the full measures equality, $\varphi = \Phi$,
leads to to the Liouville like equation for the dilaton field
\be
G^{\mu \nu} \partial_\mu \Phi  \partial_\nu \Phi - e^{-4 \Phi} = 0    \,  .  \leq{lle}
\ee
For $D=2$ it turns to the real Liouville equation.

Note that there are some considerable differences between these approaches.
Their result has been obtained in the leading order perturbation theory in
powers of the curvature, while our result is non perturbative. Their result is
a consequence of quantum one loop computation, while our result is classical.

\section{Conclusions}

In this paper, we considered classical theory of the bosonic string
propagating in the nontrivial background. In particular, we are interested in
the space-time geometry felt by the string.

In Sec 2, we investigated geometry of the surface embedded into space-time
with torsion and nonmetricity. The breaking of space-time metric postulate
produces two forms of SFF, ${}^\circ b_{i \alpha \beta}$ \eq{sff} and
${}^\circ {\bar b}_{i \alpha \beta}$  \eq{sff2}, and consequently, two forms
of MEC, whose difference is proportional to nonmetricity. We cleared the
meaning of MEC in Minkowski space-time ${}^\circ H^i$ (see \eq{aec} and
\eq{aec1}), and introduced the concept of DMEC ${}^\star H^i$ \eq{dmc} as
orthogonal projection of the mean torsion. In order to find geometrical
interpretation for the field equations, we defined C-duality which maps MEC to
DMEC. The torsion changes the equation of embedded surfaces. Instead of the
usual minimal surface ${}^\circ H_i = 0$, we introduced C-dual (antidual)
surface defined by the self-duality (self-antiduality) conditions, ${}^\circ H_i
= \pm {}^\ast H_i$.

Then we considered the equations of motion \eq{lJ}-\eq{lF}, which have been
derived in ref. \cite{BS} using Hamiltonian approach, and independently in
Sec. 3 using Lagrangian approach. With the help of the general decomposition
of the space-time connection, we have concluded that the stringy space-time
has nontrivial torsion and nonmetricity, originating from the antisymmetric
field $B_{\mu \nu}$ and dilaton fields $\Phi$. We obtained their explicit
expressions \eq{T} and \eq{mp}.

In sec. 4, we clarified the space-time geometry dependence on the background fields. In the
presence of the metric tensor $G_{\mu \nu}$, the space-time is of the Riemann type, while the
world-sheet is a minimal surface. The inclusion of the target space antisymmetric field $B_{\mu
\nu}$ produces Riemann-Cartan space-time and the world-sheet becomes C-dual surface \eq{sdrc}. In
both cases the intrinsic metric and connection are equal to the induced ones, up to conformal
transformation.

The appearance of the dilaton field $\Phi$ broke the compatibility between the
space-time metric tensor and stringy connection. It also broke conformal
invariance, introducing new component of the intrinsic metric tensor $F$, and
consequently, the new equation of motion $[i^F]$. This new field equation
allows us to calculate the induced world-sheet curvature \eq{itdc} as a
function of the dilaton field.

When all three background fields $G_{\mu \nu}$, $B_{\mu \nu}$ and $\Phi$ are
present, the string feels the complete stringy space-time and the world-sheet
becomes stringy C-dual surface \eq{ssde}. The theory looses conformal
invariance, and the relation between the intrinsic and induced variables is
fixed \eq{gg}. The corresponding factor is the length of the world-sheet
projection of the gradient of the dilaton field.

In Sec. 5, we constructed the integration measure for the theories with
nonmetricity. In fact, the stringy Weyl vector is a gradient of the scalar
field $\varphi$, necessary to make the integration measure invariant under
parallel transport. We discussed the connection between our measure and that
of the papers  \cite{CFMP,BNS}, in spite of their quite different origin.
Their result is quantum and perturbative while our is classical but non
perturbative. In particular, our scalar field $\varphi$, defined in \eq{sf},
is different from their $\Phi$, but has the same position in the expression
for the measure. These two measures are equal for $\varphi = \Phi$, which is
the condition on the dilaton field $\Phi$ in the form of the Liouville like
equation \eq{lle}.

\appendix 

\section{World-sheet geometry}

It is useful to parameterize the intrinsic world-sheet metric tensor
$g_{\alpha \beta }$, with the light-cone variables  $(h^+,h^-, F)$ (see the
papers \cite{BS,BPS,SM})
\be
g_{\alpha \beta} =e^{2F} {\hat g}_{\alpha \beta}=
\fr{1}{2}e^{2F}\pmatrix{ -2h^-h^+    &  h^-+h^+ \cr
           h^-+h^+    &  -2      \cr }\,  .               \leq{g}
\ee
The world-sheet interval
\be
ds^2 = g_{\alpha \beta} d \xi^\alpha d \xi^\beta = 2 d \xi^+ d
\xi^-  \,   ,
\ee
can be expressed in terms of the variables
\be
d \xi^\pm = { \pm 1 \over \sqrt{2}} e^F ( d \xi^1 - h^\pm d \xi^0)
= e^F d {\hat \xi}^\pm  = e^\pm{}_\alpha  d \xi^\alpha \,    .
\ee
The quantities $ e^\pm{}_\alpha$ define the light-cone one form basis,
$\theta^\pm = e^\pm{}_\alpha d \xi^\alpha$, and its inverse define the tangent
vector basis, $e_\pm = e_\pm{}^\alpha \partial_\alpha = \partial_\pm$. We will
use the relations
\be
\eta^{ab} e_a{}^\alpha e_b{}^\beta = e_+{}^\alpha e_-{}^\beta +
e_-{}^\alpha e_+{}^\beta = g^{\alpha \beta} \,  , \qquad
\varepsilon^{ab} e_a{}^\alpha e_b{}^\beta = e_+{}^\alpha
e_-{}^\beta - e_-{}^\alpha e_+{}^\beta = {\varepsilon^{\alpha
\beta} \over \sqrt{-g}} \,  , \leq{mtast}
\ee
where $a,b \in \{+,-\} $.

In the tangent basis notation, the components of the arbitrary vector
$V_\alpha$ have the form
\be
V_{\pm}=e^{-F} {\hat V}_\pm =e_{\pm}{}^\alpha V_\alpha =
{\sqrt{2} e^{-F} \over h^- -h^+} (V_0+h^{\mp}V_1)\,  .     \leq{vec}
\ee

In this notation, the Laplace operator becomes $ \stackrel{\con}{\Delta} =
e^{-2F} {\hat \Delta}$ where ${\hat \Delta}=-2 {\hat \nabla}_\pm {\hat
\partial}_\mp$. We also use the relation
\be
\sqrt{-g}  \stackrel{\con}{R}^{(2)} =\sqrt{-{\hat g}}({\hat R}^{(2)} +2 {\hat \Delta}F) \,   .        \leq{RR}
\ee

\section{Space-time geometry and world-sheet as embedded surface}

In this appendix, we introduce some notations and define the properties of $D$
dimensional space time $M_D$. We also present space-time and world sheet
classification, which depends on the background fields.

$\blacktriangleright$ In the {\it affine} space-time, $A_D \equiv (M_D,
{}^\circ \Gamma, G)$, the linear connection
\be
{}^\circ \Gamma_{\mu ,\rho \sigma} = \Gamma_{\mu ,\rho \sigma} + {}^\circ K_{\mu \rho \sigma} +
\frac{1}{2} {}^\circ Q_{\{ \mu \rho \sigma \} }   \,   ,     \leq{cdea}
\ee
can be expressed in terms of Christoffel one, contortion and nonmetricity. The
{\it Christoffel} connection
\be
\Gamma^\rho_{\nu \mu}={1 \over 2} G^{\rho \sigma}(\partial_\nu G_{\mu \sigma}
+ \partial_\mu G_{\nu \sigma}- \partial_\sigma G_{\nu \mu})   \,  ,        \leq{coG}
\ee
depends only on the space time metric tensor $G_{\mu \nu}$. The contortion is
a function of the torsion
\be
{}^\circ K_{\mu \rho \sigma} = \frac{1}{2} {}^\circ T_{ \{ \sigma \mu \rho \} } =
\frac{1}{2} ({}^\circ T_{\rho \sigma \mu} + {}^\circ T_{\mu \rho \sigma} - {}^\circ T_{\sigma \mu \rho})    \,  ,  \leq{Ka}
\ee
which itself is defined as
\be
{}^\circ T^\rho_{\mu \nu} ={}^\circ \Gamma^\rho_{\mu \nu} - {}^\circ \Gamma^\rho_{\nu \mu}
\,  .
\ee
The nonmetricity tensor is
\be
{}^\circ Q_{\mu \rho \sigma}=- {}^\circ D_\mu G_{\rho \sigma}  \,  .  \leq{nma}
\ee

\FIGURE{\epsfig{file=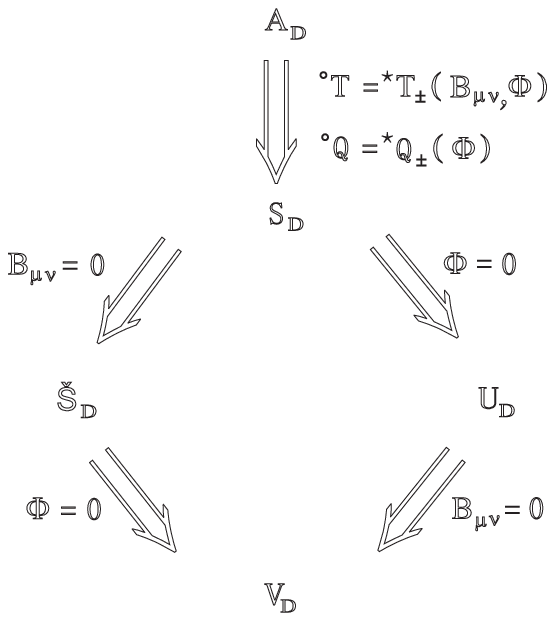,width=9cm}
        \caption[Example]{{\it Classification of space-time.} $A_D$ {\it is affine space-time},
        $S_D$ {\it is stringy space-time}, \v S${}_D$ {\it is stringy torsion free space-time},
        $U_D$ {\it is stringy Riemann-Cartan space-time and} $V_D$ {\it is Riemann space-time.}}%
    \label{stcl}}

All covariant derivatives are defined in the standard way
\be
{}^\circ D_{\mu} V_{\nu} = \partial_\mu V_{\nu} - {}^\circ \Gamma^\rho_{\nu \mu}  V_{\rho}  \,  ,  \qquad
{}^\circ D_{ \mu} V^{\nu} = \partial_\mu V^{\nu} + {}^\circ \Gamma^\nu_{\rho \mu}  V^{\rho}   . \leq{cd}
\ee

The world-sheet is {\bf affine C-dual (antidual) surface}
\be
{}^\circ H^i = \pm  {}^\ast H^i  \,    \qquad \Longleftrightarrow  \qquad
\left( G^{\alpha \beta} \mp {\varepsilon^{\alpha \beta} \over
\sqrt{-G_2}} \right) {}^\circ \! b^i_{\beta \alpha} = 0  \,      .  \leq{sdra}
\ee

$\blacktriangleright$ In the {\it Riemann} space-time, $V_D \equiv (M_D,
\Gamma, G)$, the torsion and nonmetricity vanish ${}^\circ T^\mu_{\rho
\sigma}=0$, ${}^\circ Q^\mu_{\rho \sigma}=0$, and the connection is just the
Christoffel one ${}^\circ \Gamma^\rho_{\nu \mu} = \Gamma^\rho_{\nu \mu}$.

The world-sheet is {\bf minimal surface}
\be
H^i \equiv \fr{1}{2}G^{\alpha \beta} b^i_{\beta \alpha} = 0 \,  .
\ee

$\blacktriangleright$ In the {\it stringy Riemann-Cartan} space-time, $U_D
\equiv (M_D, \Gamma_\pm, G)$, there are two types of the torsion
\be
{}^\circ T^\mu_{ \rho \sigma} \to T^\mu_{\pm \rho \sigma}=
\pm 2 B^\mu_{ \rho \sigma}   \,   ,
\ee
where the new term is the field strength of the antisymmetric tensor
\be
B_{\mu \nu \rho}= \partial_\mu B_{\nu \rho} + \partial_\nu B_{\rho \mu} + \partial_\rho B_{\mu \nu}=
D_\mu B_{\nu \rho} + D_\nu B_{\rho \mu} + D_\rho B_{\mu \nu}   \,  .       \leq{fsB}
\ee

Consequently, there are two types of the connection
\be
{}^\circ \Gamma^\rho_{\nu \mu} \to  \Gamma^\rho_{\pm \nu \mu}= \Gamma^\rho_{\nu \mu} \pm B^\rho_{\nu \mu}    \,  ,  \leq{coB}
\ee
which can be expressed in terms of the variables $G_{\pm \mu \nu} \equiv \pm 2
\Pi_{\pm \mu \nu} = G_{\mu \nu} \pm 2 B_{\mu \nu}$
\be
\Gamma_{\pm \rho, \nu \mu}= \frac{1}{2} (\partial_\nu G_{\pm \mu \rho}
+\partial_\mu G_{\pm \rho \nu}-\partial_\rho G_{\pm \mu \nu}) \,  ,       \leq{coP}
\ee
in the similar way as $\Gamma_{\rho, \nu \mu}$ can be expressed in terms of
$G_{\mu \nu}$. The nonmetricity in the stringy Riemann-Cartan space-time
vanishes, ${}^\circ Q_{\mu \rho \sigma}=0$.

The world-sheet is {\bf C-dual (antidual) surface}
\be
H_i = \pm {}^\ast H_{\mp i} \,  ,  \qquad \Longleftrightarrow  \qquad
\left( G^{\alpha \beta} \mp { \varepsilon^{\alpha \beta} \over \sqrt{-G_2}} \right)
b_{\mp i \beta \alpha} = 0    \,  .   \leq{sdrca}
\ee

$\blacktriangleright$ In the {\it stringy torsion free} space-time, \v S${}_D
\equiv (M_D, {}^\star \Gamma, G)$, the torsion vanishes ${}^\circ T^\mu_{ \rho
\sigma} = 0$, while the nonmetricity and connection obtain the forms
\be
{}^\circ Q_{ \mu \rho \sigma}  \to {}^\star Q_{ \mu \rho \sigma} \equiv  -{}^\star D_\mu G_{\rho \sigma} =
{1 \over a^2} D_\mu (a_\rho a_\sigma )       \,   ,   \leq{mptfa}
\ee
and
\be
{}^\circ \Gamma^\rho_{ \nu \mu} \to {}^\star \Gamma^\rho_{ \nu \mu}= \Gamma^\rho_{\nu \mu} +{a^\rho \over a^2} D_\mu a_\nu \,  .  \leq{ctfa}
\ee

The world-sheet is {\bf stringy minimal surface}
\be
{}^\star H_i  \equiv \fr{1}{2} G^{\alpha \beta} \, {}^\star b_{i \beta \alpha} = 0  \,   .    \leq{stfea}
\ee

$\blacktriangleright$ Finally, in the {\it stringy} space-time, $S_D \equiv
(M_D, {}^\star \Gamma_\pm, G)$, both the torsion and the nonmetricity survive
in the following forms
\be
{}^\circ T^\rho_{\mu \nu}  \to {}^\star T_\pm{}^\rho_{\mu \nu}  =
\pm 2 P^{T \rho}{}_\sigma  B^\sigma_{\mu \nu} = \pm 2 G^{(D-1) \rho}{}_\sigma B^\sigma_{\mu \nu}        \,  ,   \leq{Ta}
\ee
\be
{}^\circ Q_{\mu \rho \sigma}  \to {}^\star Q_{\pm \mu \rho \sigma} \equiv  -{}^\star D_{\pm \mu} G_{\rho \sigma} =
{1 \over a^2} D_{\pm \mu} (a_\rho a_\sigma )       \,   ,   \leq{mpa}
\ee
with the contribution of all three background fields $G_{\mu \nu}$, $B_{\mu
\nu}$ and $\Phi$. The stringy connection
\be
{}^\circ \Gamma^\rho_{\nu \mu}  \to {}^\star \Gamma^\rho_{\pm \nu \mu}= \Gamma^\rho_{\pm \nu \mu} +{a^\rho \over a^2}
D_{\pm \mu} a_\nu = P^{T \rho}{}_\sigma \Gamma^\sigma_{\pm \nu \mu} +
{a^\rho \over a^2} \partial_\mu a_\nu  \,  ,     \leq{coFi}
\ee
has the following symmetry
\be
{}^\star \Gamma^\rho_{\pm \nu \mu} = {}^\star \Gamma^\rho_{\mp \mu \nu}  \,    .   \leq{sp}
\ee

Covariant derivatives of the vector field $a_\mu$, have a properties $D_{\pm
\mu} a_\nu = D_{\mp \nu} a_\mu$ and ${}^\star D_{\pm \mu} a_\nu =0$.

The world-sheet is {\bf stringy C-dual (antidual) surface}
\be
{}^\star H_i = \pm {}^\ast H_{\mp i} \,  , \qquad  \Longleftrightarrow   \qquad
\left( G^{\alpha \beta} \mp { \varepsilon^{\alpha \beta} \over
\sqrt{-G_2}} \right) {}^\star b_{\mp i \beta \alpha} = 0   \,  .            \leq{ssdea}
\ee

\end{document}